\documentclass[preprint]{vgtc}               

\ifpdf
  \pdfoutput=1\relax                   
  \usepackage{graphicx}                
  \DeclareGraphicsExtensions{.pdf,.png,.jpg,.jpeg} 
\else
  \usepackage{graphicx}                
  \DeclareGraphicsExtensions{.eps}     
\fi%

\graphicspath{{figures/}{pictures/}{images/}{./}} 

\usepackage{microtype}                 
\PassOptionsToPackage{warn}{textcomp}  
\usepackage{textcomp}                  
\usepackage{mathptmx}                  
\usepackage{times}                     
\usepackage{cite}                      
\usepackage{tabu}                      
\usepackage{booktabs}
\usepackage{soul}
\usepackage{amsmath}
\usepackage{multirow}
\usepackage{tabularx}
\usepackage{booktabs}
\usepackage{gensymb}
\usepackage{tabularray}
\usepackage[table,xcdraw]{xcolor}
\usepackage{tabularx, multirow}
\usepackage{wrapfig}

\usepackage{tabularray}
\UseTblrLibrary{booktabs}
\usepackage{hyperref}
\usepackage{booktabs}
\usepackage{float}

\onlineid{0}

\vgtccategory{Research}

\vgtcinsertpkg

\preprinttext{This is the author's version of the article, to appear in a future issue of IEEE TVCG.}

\title{Revisiting Performance Models of Distal Pointing Tasks in Virtual Reality}

\author{Logan Lane\thanks{e-mail: logantl@vt.edu}\\ %
        \scriptsize Virginia Tech %
\and Feiyu Lu\thanks{e-mail: feiyulu@vt.edu}\\ %
     \scriptsize Virginia Tech %
 \and Shakiba Davari \thanks{e-mail: sdavari@vt.edu}\\ %
    \scriptsize Virginia Tech %
\and Robert J. Teather \thanks{e-mail: Rob.Teather@carleton.ca}\\ %
    \scriptsize Carleton University %
 \and Doug A. Bowman\thanks{e-mail: dbowman@vt.edu}\\ %
 \scriptsize Virginia Tech %
}

\teaser{
  \centering
  \includegraphics[width=\linewidth]{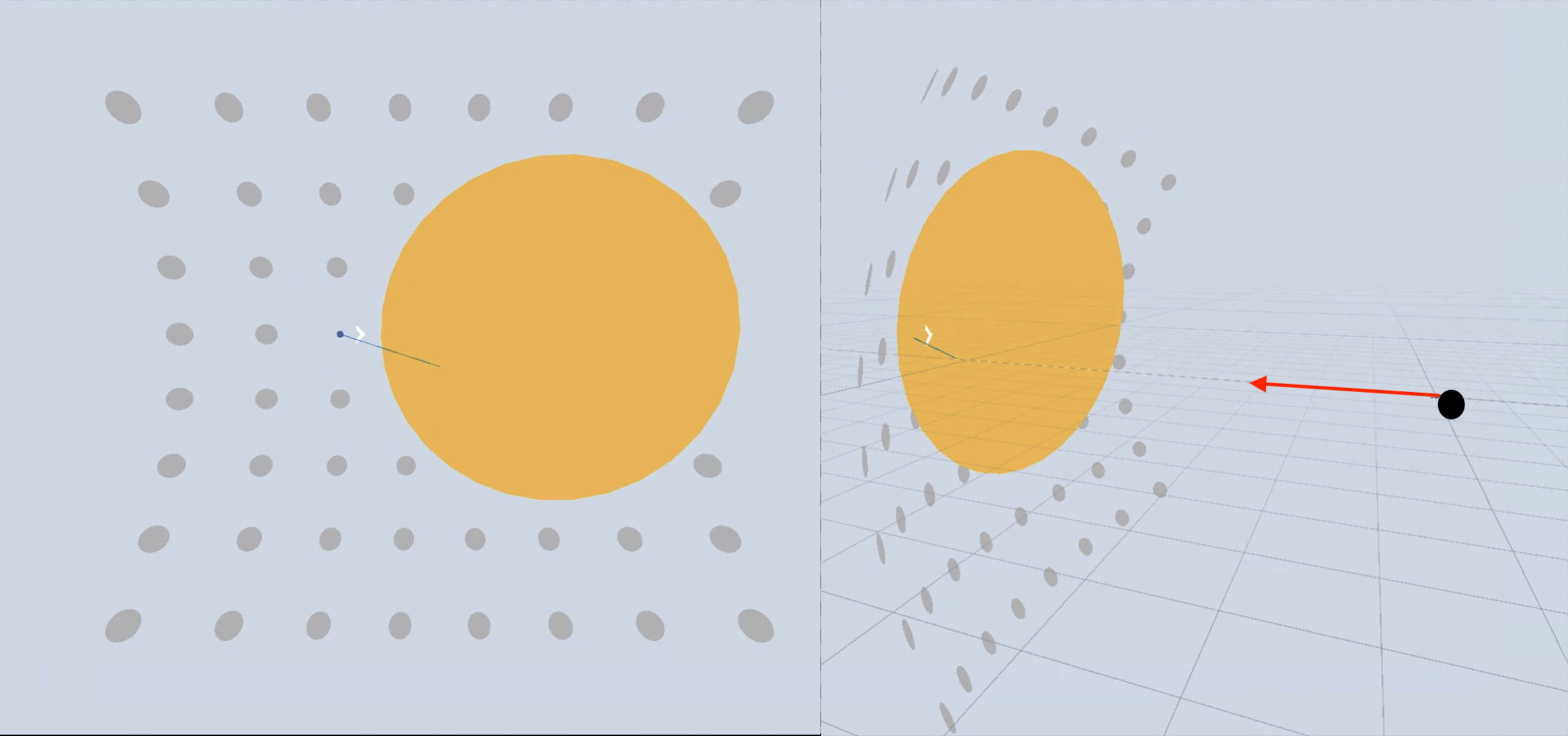}
  \caption{The proposed replacement for the ISO 9241-411 standard selection task. The starting object is represented by the blue sphere and the target is represented by the yellow sphere. The left portion of the figure shows the task from the participant's point of view. The right portion shows how the task looks from a third-person point of view. The black dot represents the location of the participant and the red arrow indicates the direction they were facing.}
  \label{fig:teaser}
}

\abstract{Performance models of interaction, such as Fitts’ law, are important tools for predicting and explaining human motor performance and for designing high-performance user interfaces. Extensive prior work has proposed such models for the 3D interaction task of distal pointing, in which the user points their hand or a device at a distant target in order to select it. However, there is no consensus on how to compute the index of difficulty for distal pointing tasks. We present a preliminary study suggesting that existing models may not be sufficient to model distal pointing performance with current virtual reality technologies. Based on these results, we hypothesized that both the form of the model and the standard method for collecting empirical data for pointing tasks might need to change in order to achieve a more accurate and valid distal pointing model. In our main study, we used a new methodology to collect distal pointing data and evaluated traditional models, purely ballistic models, and two-part models. Ultimately, we found that the best model used a simple Fitts'-law-style index of difficulty with angular measures of amplitude and width. 
} 

\CCScatlist{
  \CCScatTwelve{Human-centered computing}{Human Computer interaction (HCI)}{Interaction techniques}{Pointing};
  \CCScatTwelve{Human-centered computing}{Human Computer interaction (HCI)}{Interaction paradigms}{Virtual reality};
}

\begin{document}


\firstsection{Introduction}

\maketitle





Distal pointing, in which the user points their hand or a device at a distant target in order to select it, is a key component of interactions in 3D user interfaces. Ray casting, an interaction technique based on distal pointing,  is often used as a standard selection technique in today's 3D user interfaces, such as Meta's Quest Virtual Reality (VR) platform and the Magic Leap Augmented Reality (AR) Head Worn Display (HWD). Distal pointing allows users of these 3D interfaces to interact with elements of the interface from a distance.

It is beneficial to UX practitioners to be able to accurately predict the time that it will take to complete distal pointing selection tasks and other common interactions when interacting with a 3D interface. An accurate performance model would allow researchers to better understand user behavior and as a result create more efficient user interfaces. It also has the added benefit of accurately predicting untested difficulty values. Fitts' law\cite{fitts_information_1954} is the basis for many predictive models that allow UX practitioners to predict the time it will take a user to complete a given interaction in a user interface. Fitts’ law and similar models accomplish this by predicting selection time as as a function of an "index of difficulty" (ID), which in turn is a function of the target size (W) and movement amplitude (A) \cite{kabbash1995prince, schmitt1999calculation, thompson2004gain, ahlstrom2005modeling}.

Fitts’ law was originally created to model the performance of one-dimensional (1D) tapping tasks \cite{fitts_information_1954}. However, Fitts’ law has been shown to also work well with two-dimensional (2D) selection tasks \cite{mackenzie_extending_1992, mackenzie_fitts_1992, murata_empirical_1996, murata_extending_1999}. There is clear interest in extending Fitts’ law to three-dimensional (3D) selection tasks. Several attempts have been made to extend Fitts' law to work for 3D distal pointing\cite{amini2025review3dfitts}. Researchers such as Murata and Iwase\cite{murata_extending_2001}, and Kopper et al.\cite{kopper_human_2010} have attempted to extend Fitts' Law, sometimes adding parameters such as movement direction or distance to the surface on which targets lie, or replacing linear measures of size and amplitude with angular measurements, to account for differences between distal pointing and traditional 2D pointing tasks. However, it is unclear which form of model works best when modeling selection performance in a 3D user interface. Existing performance models were generated based on studies with older technology that are outdated or non-standard by the standards of today's modern VR and AR equipment. There has been little work conducted regarding modeling selection performance using modern HWDs. For example, the experiment of Kopper et al. utilized large displays to show small targets to the participants \cite{kopper_human_2010}. It is unclear whether these results generalize to HWDs.

Existing models are also not as elegant as the standard Fitts' law model. Models that have attempted to predict user selection performance in 3D user interfaces have often had to resort to some unconventional methods in order to produce a model that fits their data correctly. For example, the best performing model from Kopper et al.'s work \cite{kopper_human_2010} needed to cube the target size value and square the entire index of difficulty in order to produce a satisfactory fit, without clear justification for doing so. Thus, we suggest that it is time to revisit distal pointing models for VR. 

Modeling Fitts'-style pointing tasks is so fundamental that a standard task methodology (ISO-9241-411) has been defined for this purpose. The standard task arranges circular targets in a circular layout and has users select all the items in a reciprocal fashion (i.e., always moving from a target on one side of the circle to one on the other side). However, the standard task is not without its limitations, such as a limited range of possible ID values (especially easy ones) and trial interdependence. Due to these limitations, rethinking the existing methods for collecting aimed movement data may also be useful in improving distal pointing models.

Our work seeks to answer the following research questions. 

\begin{itemize}
    \item \textbf{RQ1: Is there a simple and elegant model that can accurately predict distal pointing performance across a wide range of realistic task difficulties?}
    \item \textbf{RQ2: How does user selection behavior change based on task difficulty, and how can that inform the design of a predictive model?}
\end{itemize}

To address these questions, we conducted a preliminary study in which we compared the predictions of existing models to actual task performance in VR using the ISO 9241-411 standard selection task. This analysis suggested that selection tasks with a very low index of difficulty are not modeled well by existing models. Because of this, we hypothesized that it might be necessary to create a new two-part model that separately models easy and regular difficulty distal pointing tasks. 

After developing a new task and evaluation methodology, we conducted a second user study to generate more data to test the idea of a two-part model alongside other model formats. Ultimately, we found that a simple model using angular measures of size and amplitude modeled the data best. Our data also allowed us to do a deeper analysis of user selection behavior for distal pointing tasks with a wide range of difficulties.

The contributions of our work are as follows:
\begin{itemize}
    \item An evaluation of different models for distal pointing in the literature
    \item Discussion and assessment of the idea that very easy distal pointing tasks might need to be modeled separately
    \item A novel, generic distal pointing assessment methodology that provides data on distal pointing tasks for a wide range of difficulties without interdependence between trials 
    \item An elegant predictive model for distal pointing tasks that accurately models performance across a wide range of difficulties
\end{itemize}

\section{Related Work}
In 1954, Paul Fitts proposed a model for modeling the time it takes a user to move reciprocally between physical plates \cite{fitts_information_1954}. Fitts found that we could accurately model the time it takes an individual to move between targets of different sizes and distances from one another using a reciprocal selection pattern when moving between targets. Fitts' law is defined as:

\begin{equation}
    MT = a + b \cdot ID, \quad \text{where} \quad ID = \log_2\left(\frac{2A}{W}\right)
\end{equation}

 where $A$ is the amplitude or distance of the target from the starting point, $W$ is the width of the target, and $a$ and $b$ are both determined via linear regression. The original model provided by Fitts has been reinterpreted by multiple researchers \cite{welford1968fundamentals, shoemaker, murata_extending_2001, cha_extended_2013}; specifically, others have reinterpreted the formula for ID. One interpretation by MacKenzie, also known as the Shannon formulation, has served as the standard for future distal pointing research \cite{mackenzie_extending_1992}. MacKenzie's reinterpretation of Fitts' law is: 

 \begin{equation}
   MT = a + b \cdot ID, \quad \text{where} \quad ID = \log_2\left(\frac{A}{W} + 1\right)
\end{equation}

This model has been used to model performance in the context of 2D user interfaces for many decades \cite{kabbash1995prince, schmitt1999calculation, thompson2004gain, ahlstrom2005modeling}. Many attempts have been made to extend this model to 3D user interfaces with varying degrees of success. This model will serve as the base for the various models used throughout this paper. Specifically, we will alter how we calculate ID.

The challenges of extending this model to work in 3D user interfaces are numerous as noted by Triantafyllidis and Li \cite{triantafyllidis_challenges_2021}. The following sections will describe previous work in modeling the performance of 3D interactions, modeling distal pointing performance, and experimental tasks for evaluating distal pointing. 

\subsection{Modeling 3D Interactions}
Goals, Operators, Methods, and Selection (GOMS) and Keystroke-Level Models (KLM) are predictive models that allow designers to predict the amount of time it would take a user to complete a given interaction. GOMS accomplishes this by defining: Goals that a user would want to accomplish when using software, Operators which are actions that a user can perform within software, Methods which are combinations of subgoals and operators that a user can perform within the application, and selection rules which are determined by the individual user to decide which methods and operators to use to accomplish a goal \cite{john1995goms, john1996using}. Several of these predictive models have been created for 2D interfaces \cite{john1995goms, john1996goms, john1996using, kieras1997guide}. In modeling the performance of 3D interactions, researchers strive to create a GOMS/KLM style predictive model that would allow designers to predict the time that it would take a user to complete a given 3D interaction. 

There has been some early work conducted in progressing towards a predictive model for 3D interactions. Namely works by Cabric et al, Ghasemi et al., and Zhou et al. began exploring the idea of extending existing GOMS and KLM style models to support interactions in AR and VR \cite{cabric_predictive_2021, ghasemi_evaluating_2023, zhou_h-goms_2023}. However, the individual interactions must first be understood and predictable on their own. There has been much work conducted in an effort to achieve this goal. Specifically, many researchers have attempted to extend Fitts' law to work with other 3D interactions besides distal pointing. Zhao et al. and Grossman and Balakrishnan both conducted studies that attempted to extend Fitts' law to direct touch interactions\cite{zhao_movement_2023, inproceedings_Grossman, cha_extended_2013}. 

In summary, there is still a significant amount of work required in modeling individual interactions before we can develop an all-encompassing, GOMS/KLM style predictive model for 3D interactions. The rest of this paper will focus exclusively on one component of these future GOMS/KLM models, modeling distal pointing performance. 

\subsection{Modeling Distal Pointing}
Several researchers have attempted to extend Fitts' law to account for distal selection tasks in a 3D interface. Murata and Iwase were among the first to apply Fitts' original model to 3D distal pointing tasks\cite{murata_extending_2001}. They found that while Fitts' law worked well for 1D and 2D tasks (as proven by Mackenzie and Buxton \cite{mackenzie_extending_1992} and Murata \cite{murata_empirical_1996} \cite{murata_extending_1999}) the conventional Fitts' model, modified by Mackenzie in 1989\cite{mackenzie_note_1989}, did not fit the data nearly as well for 3D distal pointing tasks. As a result, they proposed a new model that took into account the direction a user had to move in order to hit the target by including theta (the direction) into the formula for ID.

In the time since these early works were published, multiple researchers have attempted to use and extend these early models to model 3D distal pointing performance to varying degrees of success. Some researchers have looked at the inclination of the user interface to see if that would provide a greater data to model fit \cite{clark_extending_2020}. There have also been attempts to add target depth to existing models to see how that affected model fit \cite{wagner_fitts_2023, barrera_machuca_effect_2019, teather_pointing_2013, kopper_human_2010}. Specifically, Teather and Stuerzlinger found that modeling 3D interactions (both direct touch and distal pointing) using Fitts' law would require a change in the formulation of the 2D Fitts' law model as it did not perform well in modeling 3D selection tasks that were not presented at or parallel to the screen \cite{teather_pointing_2011}.

Researchers have also attempted to extend Fitts' law by accounting for the angular distance that a user has to move between a starting point and the target ($\alpha$) and the angular width of the target ($\omega$). Using angular measurements in Fitts' law calculations rather than linear measurements makes sense because most of the movements involved in distal pointing tasks are rotations of the elbow or wrist rather than hand translations. Kopper et al. wrote two papers that dealt with performance modeling in 3D distal pointing and selection tasks. The first of these papers provided a new $ID_{\text{ANG}}$ model for modeling performance in distal pointing tasks. \cite{kopper_human_2010, kopper_rapid_2011}.

\[ID_{\text{ANG}} = \log_2\left(\frac{\alpha}{\omega} + 1\right)\]

in which $\alpha$ is the angular distance between the center of the starting object to the center of the target, from the user's perspective, and $\omega$ is angular size of the target from the user’s perspective.

The authors concluded that this model fit their experimental data mostly well, but they noticed an exponential outlier trend occurring within the model. To remedy this, they further modified their proposed model by providing an exponent (K) to the $\omega$ value in the updated $ID_{\text{DP}}$ model and squaring the logarithm result. 

$$ID_{\text{DP}} = \left[\log_2\left(\frac{\alpha}{\omega^k} + 1\right)\right]^2$$

Through regression analysis, the researchers determined the optimal fit for the model occurred when k = 3. By providing $\omega$ with a cube exponent, they were able to fit their data to the model better. We utilize the $ID_{\text{ANG}}$ and $ID_{\text{DP}}$ models in our paper as comparison models for our own data. We use the $ID_{\text{ANG}}$ model as it is essentially an angular version of the Shannon formulation of Fitts' law \cite{mackenzie_extending_1992}. The $ID_{\text{DP}}$ model is included here as it was used in Kopper et al.'s previous work and found to fit their data well\cite{kopper_human_2010}. Similarly, we included an  $ID_{\text{ANG}^3}$ model that took the angular Fitts' law formulation from Kopper et al.'s work and cubed $\omega$, following the results from that work that found their data fit best when k (the exponent on $\omega$) equaled 3 \cite{kopper_human_2010}. The $ID_{\text{ANG}^3}$ model is defined below:

\[ID_{\text{ANG}^3} = \\log_2\left(\frac{\alpha}{\omega^3} + 1\right)\]

Throughout the past few decades, there have been many studies of various distal pointing models. The only clear consensus within the community is that Fitts' law \textbf{should} be applicable to 3D movement in the same way that it has worked for 1D and 2D movement. However, there still has not been a model that has been agreed upon by the community at large as being reliable while being simple, elegant, and explainable. Works by Holmes et al., Zeng et al., and Burno et al. seem to suggest that existing Fitts' law performance models could still model 2D targets in 3D space well, however, there still is not an agreed upon model for 3D distal pointing selections in the same manner as 1D and 2D selections \cite{zeng_fitts_2012, burno_applying_2015, holmes_using_2016, amini2025review3dfitts}.

Quite a few of the aforementioned works provide a modified version of Fitts' law that has some additional element added that allowed their data to fit better when conducting a linear regression. However, adding additional terms to a linear regression will always increase the $R^2$ of predictive models, which can be misleading\cite{rousson_-square_2007}. Finding a model that fits distal selection data well while also being intuitive and explainable is still an open question. The goal of this paper is to revisit some of the popular Fitts-style models, determine if they do model distal pointing tasks well, and if not, analyze what can be done to improve the overall fit and performance modeling.

\subsection{Methods For Collecting Aimed Movement Data}

Dating back to the original tapping experiment conducted by Fitts in 1954, there have been many methods for collecting aimed movement data. Fitts created the first of these tasks which had participants tap back and forth between two targets with varying distances and target widths \cite{fitts_information_1954}. This task would go on to serve as inspiration for future iterations of methods for collecting aimed movement data. One of these iterations was created by Mackenzie et al. where they had participants click targets with differing amplitudes and widths using either a mouse, stylus, or trackball \cite{mackenzie_comparison_nodate}. 

In 1999, the ISO 9241-9:2000 standard was created as a standardization of requirements for non-keyboard input devices \cite{ISO_2010}. This standard featured guidelines for measuring the performance of non-keyboard input devices. Douglas et al. conducted one of the first experiments that utilized the performance measure guidelines from the ISO 9241 standard and had their participants complete a very early version of the ISO standard task that is seen in distal pointing studies frequently today \cite{DouglasISO}. From there, the standard was improved and iterated on until it became the gold standard for data collection in distal pointing and Fitts' law works. 

However, the ISO standard task is not without its flaws. In section 4, we discuss the limitations of the methodology with respect to trial interdependence and inability to easily test very easy distal pointing tasks; we propose a new methodology to address these shortcomings.

\section{Preliminary Experiment}
\subsection{Goals}
\begin{figure}[ht]
    \centering
    \includegraphics[width=8cm]{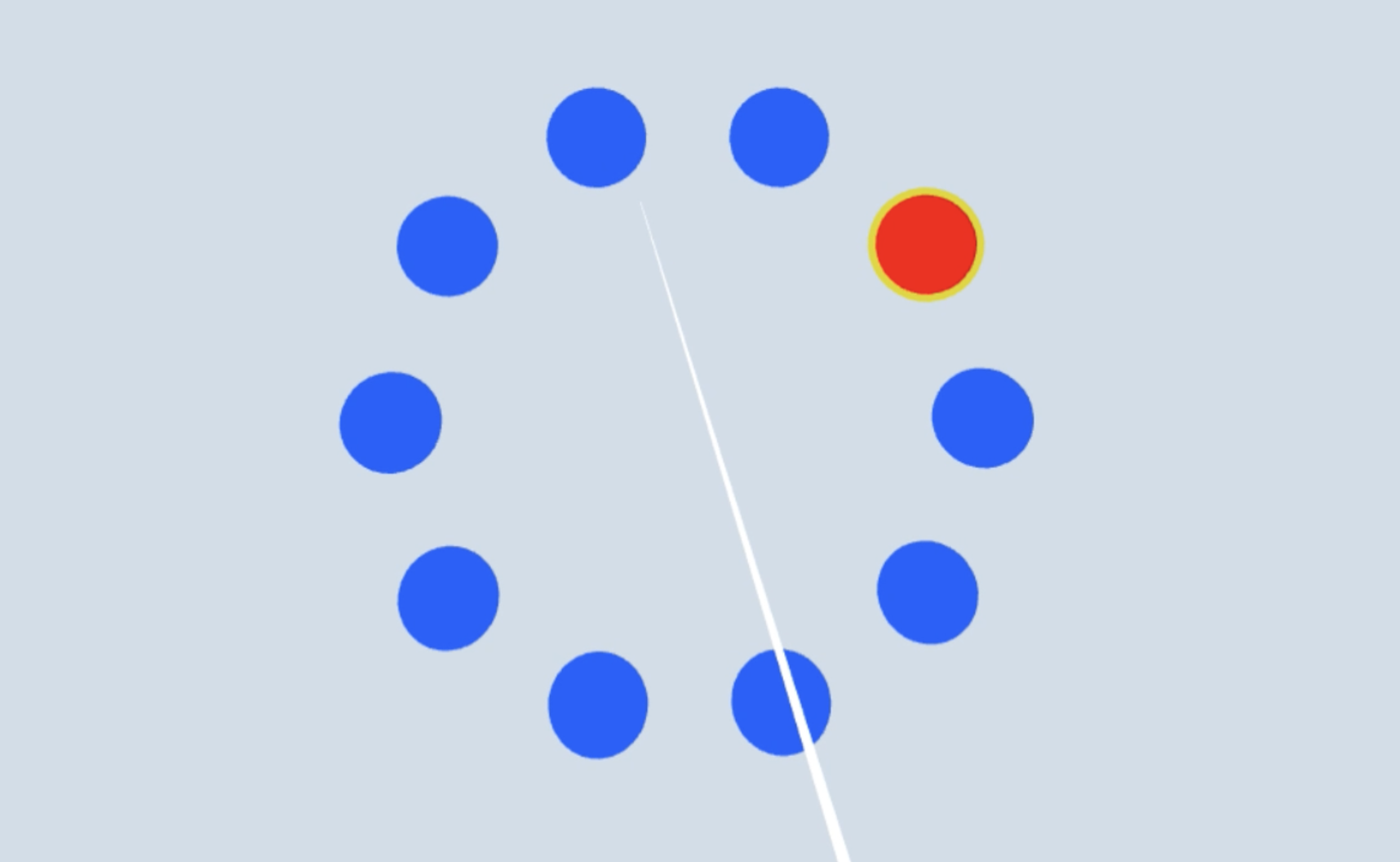}
    \caption{The selection task used in the preliminary study.}
    \label{fig:prelimtask}
\end{figure}

This preliminary experiment was conducted in order to test past 3D distal pointing models using modern HWDs that feature higher fidelity displays and tracking systems compared to the systems used in prior distal pointing work. In addition, because our prior experience led us to believe that the $ID_{\text{DP}}$ model \cite{kopper_human_2010} had poor performance on very easy distal pointing tasks, such as those found in many AR/VR menu interfaces, we wanted to gather data on such tasks. To accomplish this, the selection task utilized in the study recreated the ISO 9241-411 standard selection task as shown in Figure \ref{fig:prelimtask}. By utilizing the ISO standard selection task, the results from our study can be compared to other previous distal pointing work that also utilizes the ISO standard selection task.

\subsection{Experimental Design}
This within-subjects study had two independent variables: $\alpha$ (20, 40, 60 degrees) and $\omega$ (1, 2, 3, 5, 10, 15 degrees). Time (in seconds) was the only dependent variable in the preliminary study. In total, there were 18 possible combinations of the two independent variables. It should be noted that the easiest combination of $\alpha$ and $\omega$ ($\alpha$ = 20 and $\omega$ = 15) would provide an ID of 0.00007 using Kopper et al.'s formulation for ID, which is significantly lower than the ID values tested in their previous work \cite{kopper_human_2010}.

\subsection{Apparatus}
Participants used a Meta Quest 2 HWD, and one of two Meta Touch Controllers (participants only used the respective controller for their dominant hand.) The headset has an 1832 x 1920 resolution per eye with a refresh rate of up to 90 Hz.\footnote{https://www.meta.com/quest/products/quest-2/tech-specs/} In addition, we also used a PC running Windows 10 with an i9-12900K processor, RTX 3070 Ti graphics card, and 32GB of DDR5 RAM to run the VR experience through the Unity 2020.3.17f1 editor. 
\subsection{Task}
Participants were tasked with selecting 10 spherical targets that were arranged in a circular pattern according to the ISO 9241-411 standard selection task. The targets were placed 2 meters from the participants. The participant's controller had a white ray extending from it that they would use to select each of the targets by intersecting the ray with one of the targets. Once participants pointed the ray at a target, it was highlighted with a yellow circle indicating the ability to select the target by pulling the trigger on the controller. Participants would select the initial target highlighted in red and then select the red target opposite to the previously selected one until all targets were selected. The direction of the next target was always shown to the participant by way of a white arrow that pointed in the direction of the next target after the previous target was selected. Participants completed 54 sets of trials throughout the study with each set having 10 individual trials. In this case, a trial is the selection of an individual target in the circular arrangement of 10 targets. Each trial within a set featured 10 targets with the same $\alpha$ and $\omega$ values. Between sets, $\alpha$ and $\omega$ varied with participants seeing each of the 18 unique combinations of $\alpha$ and $\omega$ 3 times. In total, 7560 trial data points were collected across 14 study sessions.
\subsection{Procedure}
The study was approved by our local ethics board. Upon arriving to the testing area, participants were welcomed into our lab. Prior to the study, participants were emailed a copy of the consent document and asked to review it prior to participating in the study. Participants were then asked to sign the consent form acknowledging their participation in our study. 

Participants then completed the pre-study questionnaire where they provided their gender, occupation, age, and dominant hand. Participants also self-reported their fatigue level as well as their experience with VR on a Likert scale ranging from 1 to 5.  

Participants were then shown a brief presentation with a video that demonstrated to participants how to perform the selection task. Participants were then introduced to the Meta Quest 2 and the Meta Touch Controller that they would be using to complete the study. Participants were shown how to adjust the strap on the HWD so that it would fit comfortably on their head. Participants were also shown the trigger on the back of the touch controller which is how they would select the targets. Prior to the study beginning, participants completed three sets of practice trials. Participants completed a set of trials with a random easy ID, a random medium ID, and a random hard ID. After completing these practice trials successfully, participants were free to begin the study when they felt comfortable to do so. Participants then completed the 54 sets of trials in a completely random order that was generated by the Unity editor on launch. Any failed trials by the participants were given to the participants again at the end of the study. There were no additional questionnaires or interviews conducted at the conclusion of the study. 

\subsection{Participants}
We recruited 14 participants (11 Male, 3 Female) from various Human-Computer Interaction and computer science email lists. 12 of our participants self-reported as being predominantly right-handed, 1 participant self-reported as being predominantly left handed, and 1 participant self-reported as being ambidextrous. The ambidextrous participant was free to choose the hand they preferred, however, they were asked to be consistent with the hand they chose. Participants had an average age of 24.28 years old, with our youngest participant being 19 years old and our oldest participant being 38 years old. 
\subsection{Results}

\begin{table*}[ht!]
\centering
\begin{tblr}{
  width = \linewidth,
  colspec = {|X[c]|X[c]|X[c]|X[c]|X[c]|X[c]|X[c]|X[c]|},
  row{1} = {font=\bfseries},
  row{2-Z} = {font=\normalfont},
  row{3} = {font=\bfseries},
  row{9} = {font=\bfseries},
  hlines,
  vlines,
}
L-R           & BreakPoint              & \SetCell[c=3]{c} Model-L    &   &   & \SetCell[c=3]{c} Model-R    &   &   \\
              &                         & \( R^2 \)      & Intercept      & Slope             & \( R^2 \)      & Intercept       & Slope           \\
3-15          & 0.0006                  & 93.65\%        & 0.3791         & 277.1991          & 83.05\%        & 0.7019          & 0.0275          \\
4-14          & 0.0008                  & 15.24\%        & 0.4216         & 83.6293           & 86.26\%        & 0.7314          & 0.0262          \\
5-13          & 0.0032                  & 29.02\%        & 0.4443         & 30.2779           & 87.86\%        & 0.7546          & 0.0253          \\
6-12          & 0.0071                  & 66.05\%        & 0.4466         & 27.4405           & 87.99\%        & 0.7703          & 0.0246          \\
7-11          & 0.0458                  & 10.06\%        & 0.4950         & 1.6225            & 91.56\%        & 0.8049          & 0.0232          \\
8-10          & 0.16                  & 45.06\%        & 0.4976         & 1.2237            & 92.33\%        & 0.8266          & 0.0223          \\
9-9           & 0.32                  & 72.28\%        & 0.5009         & 1.0416            & 91.62\%        & 0.8294          & 0.0222          \\
10-8          & 0.64                  & 41.99\%        & 0.5268         & 0.4090            & 94.99\%        & 0.8736          & 0.0204          \\
11-7          & 1.72                  & 57.58\%        & 0.5433         & 0.2311            & 94.21\%        & 0.8779          & 0.0202          \\
12-6          & 2.85                  & 67.50\%        & 0.5536         & 0.1694            & 93.90\%        & 0.8614          & 0.0208          \\
13-5          & 3.27                  & 64.50\%        & 0.5619         & 0.1310            & 92.73\%        & 0.9075          & 0.0191          \\
14-4          & 6.68                  & 60.88\%        & 0.5860         & 0.0815            & 89.01\%        & 0.9558          & 0.0174          \\
15-3          & 9.53                  & 71.03\%        & 0.5948         & 0.0690            & 98.25\%        & 0.7125          & 0.0255          \\
\end{tblr}
\caption{Table that shows attempts made at finding a breakpoint for easy and difficult levels of ID for the \( ID_{\text{DP}} \) model.}
\label{BreakpointTableIDDP}
\end{table*}

\begin{table*}[ht!]
\centering
\begin{tblr}{
  width = \linewidth,
  colspec = {|X[c]|X[c]|X[c]|X[c]|X[c]|X[c]|X[c]|X[c]|},
  row{1} = {font=\bfseries},
  row{2-Z} = {font=\normalfont},
  row{5} = {font = \bfseries},
  hlines,
  vlines,
}
L-R           & BreakPoint              & \SetCell[c=3]{c} Model-L    &   &   & \SetCell[c=3]{c} Model-R    &   &   \\
              &                         & \( R^2 \)      & Intercept      & Slope             & \( R^2 \)      & Intercept       & Slope           \\
3-11          & 1.8745                  & 96.12\%        & 0.1967         & 0.1495            & 97.44\%        & -0.2084         & 0.2966          \\
4-10          & 2.3219                  & 98.69\%        & 0.1967         & 0.1495            & 97.54\%        & -0.2583         & 0.3073          \\
5-9           & 2.8074                  & 99.15\%        & 0.1780         & 0.1611            & 97.32\%        & -0.2846         & 0.3127          \\
6-8           & 2.9386                  & 98.79\%        & 0.1607         & 0.1716            & 97.26\%        & -0.3338         & 0.3226          \\
7-7           & 3.1699                  & 98.92\%        & 0.1683         & 0.1673            & 96.45\%        & -0.3338         & 0.3226          \\
8-6           & 3.4594                  & 94.31\%        & 0.1158         & 0.1953            & 98.30\%        & -0.4782         & 0.3503          \\
9-5           & 3.7004                  & 95.40\%        & 0.1225         & 0.1919            & 97.57\%        & -0.4841         & 0.3514          \\
10-4          & 3.8413                  & 96.17\%        & 0.1109         & 0.1975            & 97.00\%        & -0.6546         & 0.3830          \\
11-3          & 4.3923                  & 96.48\%        & 0.0819         & 0.2104            & 98.44\%        & -1.0617         & 0.4561          \\
\end{tblr}
\caption{Table that shows attempts made at finding a breakpoint for easy and difficult levels of ID for the \( ID_{\text{ANG}} \) model.}
\label{PreliminaryIDANGBreakpointTable}
\end{table*}

Throughout both the preliminary study and main study, the fit of data for a particular model was determined by calculating the grand average time it took all participants to complete a trial with a certain combination of alpha and omega values, and then performing a linear regression on the 18 resulting averages. From there, the coefficient of determination ($R^2$) value was used to determine how well the data fit a particular model.

\begin{figure}[ht]
    \centering
    \includegraphics[width=8cm]{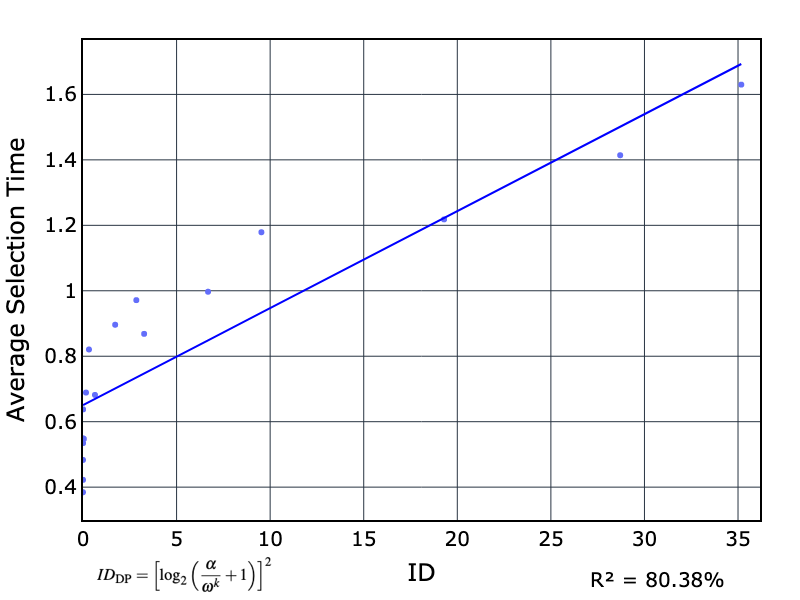}
    \caption{Linear regression plot using the data from the preliminary study for the $ID_{\text{DP}}$ model. }
    \label{fig:IDDPPrelimStudy}
\end{figure}

We used data gleaned from our preliminary study and attempted to fit the data to existing distal pointing models. We began by attempting to fit our data to the preferred $ID_{\text{DP}}$ model by Kopper et al. \cite{kopper_human_2010}. We found that our data was modeled fairly well with $R^2$ = 84.79. However, we noticed an interesting pattern when looking at the linear regression plot, shown in figure \ref{fig:IDDPPrelimStudy}. It appeared that the distal pointing tasks with low ID values at the left of the plot were not modeled particularly well at all. We hypothesized that selection tasks with a low ID value should be modeled separately from those tasks with a higher ID value, as we believed that participants could have been completing tasks with low ID values in a purely ballistic manner. The idea of such a two-part model is not unprecedented, with Shoemaker et al. initially proposing a similar approach as a replacement for Fitts' law in modeling user selection performance on 2D displays where they found that two-part models more accurately modeled distal pointing performance when compared to one-part models.\cite{shoemaker}. Two-part models have been tested in other works and have been shown to increase model fit when compared to one-part models\cite{JanzenTwoPart, Sindhupathiraja_2024}. Similarly, Gan \& Hoffman found that Fitts' law could be less accurate when describing movements with ID smaller than 3 \cite{gan_geometrical_1988-2}. Schuetz found that when ID is smaller than 1.4, target selection time using gaze becomes close to a constant value \cite{schuetz_explanation_2019}. We hypothesize that the same could apply to distal pointing scenarios with HWDs.

\begin{figure}[ht]
    \centering
    \includegraphics[width=8cm]{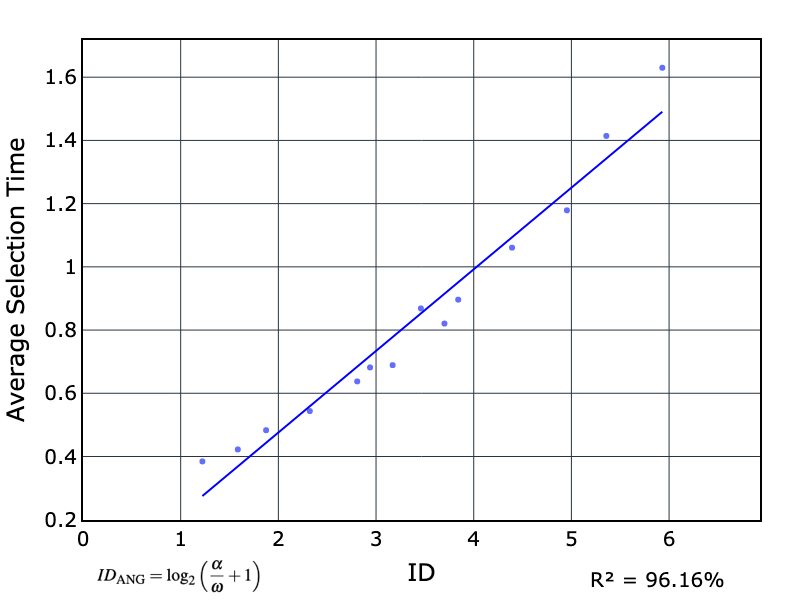}
    \caption{Linear regression plot using the data from the preliminary study for the $ID_{\text{ANG}}$ model.}
    \label{fig:regressionplotIDANG}
\end{figure}

We also attempted to fit the data from the preliminary study to both the $ID_{\text{ANG}}$ and $ID_{\text{ANG}^3}$ models. In fitting this data, we found that the $ID_{\text{ANG}}$ model modeled the preliminary study data relatively well producing an $R^2$ value of 92.4\% (see \autoref{fig:regressionplotIDANG}). However, we continued to search for an additional solution as the $ID_{\text{ANG}}$ model produced a negative intercept of -0.0406, which is undesirable as it suggests that the easiest distal pointing task would take negative time, which is, of course, impossible.

\begin{figure}[ht]
    \centering
    \includegraphics[width=8cm]{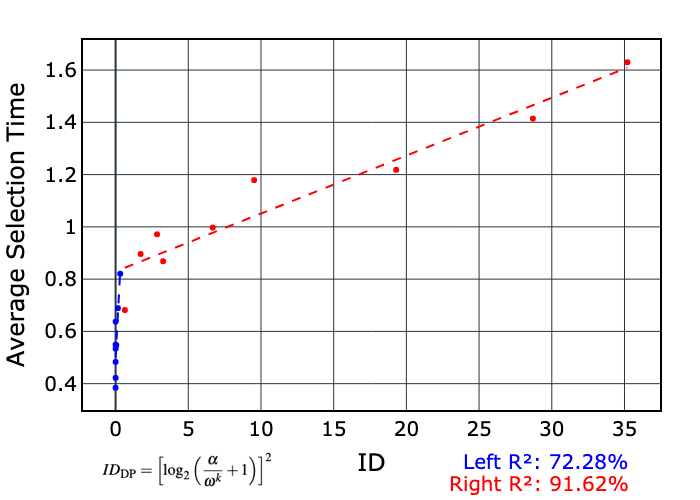}
    \caption{Linear regression plot that shows the left (blue) and right (red) regression models for ID = 0.32}
    \label{fig:splitregressionplot}
\end{figure}

The idea of modeling distal pointing tasks with a low ID value separately was explored further by creating a two-part model in which we found a ``breakpoint'' in the ID values where two separate left and right models could be created for modeling performance in distal pointing tasks. To do this, the data from the preliminary user study was used to calculate three different values of ID using the $ID_{\text{ANG}}$, $ID_{\text{ANG}^3}$, and $ID_{\text{DP}}$ models. We then ordered each of the ID values from least to greatest. We then explored two-part regression models by breaking the data into two pools at a ``breakpoint'' (i.e. an ID breakpoint of 0.16 would mean that the data associated with all IDs less than 0.16 inclusive would be one data pool while the data for all other ID values would be in another pool). The leftmost breakpoint was the third smallest ID value, while the rightmost breakpoint was the third largest ID value, in order to have enough data to perform the linear regression. Table \ref{BreakpointTableIDDP} shows the results of these models using $ID_{\text{DP}}$. The L-R column represents the progression through the distinct ID values, the BreakPoint column is the selected breakpoint for that particular analysis, and the Model-L and Model-R columns show the results for the linear regression calculations. As Table \ref{BreakpointTableIDDP} shows, the 3-15 and 9-9 splits of the data gave us the best results, though neither was particularly good at fitting both parts of the data. Figure \ref{fig:splitregressionplot} shows the linear regression plot for the 9-9 row in Table \ref{BreakpointTableIDDP}.

\begin{figure}[ht]
    \centering
    \includegraphics[width=8cm]{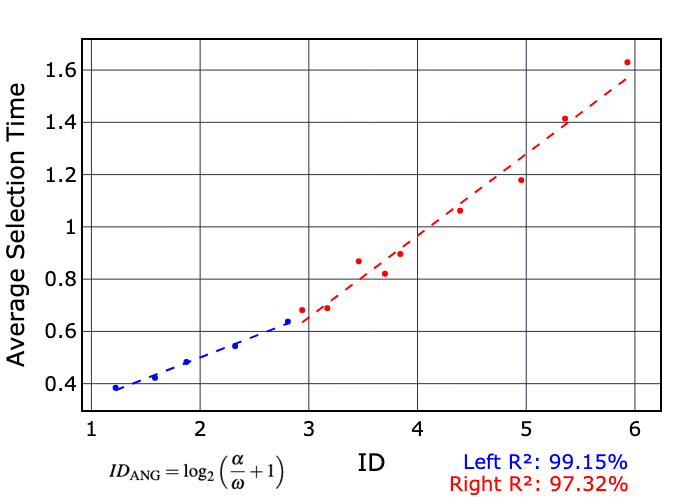}
    \caption{Linear regression plot that shows the left (blue) and right (red) regression models for ID = 2.807, the best fitting two-part model from the preliminary data.}
    \label{fig:splitregressionplotIDANG}
\end{figure}

Following the same process as above, we further evaluated the two-part model concept by utilizing the $ID_{\text{ANG}}$ formula for ID calculations. These results are shown in Table \ref{PreliminaryIDANGBreakpointTable}. We found that the two-part model with the best fit was for the 5-9 split, where the left model had an $R^2$ value of 99\% and the right model had an $R^2$ value of 97\%. The plot for this particular model is shown in Figure \ref{fig:splitregressionplotIDANG}. This two-part model granted a higher $R^2$ value compared to the one-part model which had an $R^2$ value of 96.16\%. The plot for the one-part model is shown in Figure \ref{fig:regressionplotIDANG}.

This finding prompted us to conduct a second, more comprehensive study in order to evaluate whether a two-part model would more accurately model user distal selection performance as well as further explore if the $ID_{\text{ANG}}$ model alone would be sufficient to model user performance. In this second study, a new distal pointing methodology was utilized that would allow more varied ID values to be produced so the two-part model could be tested on more unique and varied data.

\section{A New Distal Pointing Method For Collecting Aimed Movement Data}
\subsection{Limitations of ISO 9241-411}
The ISO 9241-411 standard selection task has been used in a substantial number of distal pointing research papers over the past few decades \cite{DouglasISO, kopper_human_2010, teather_pointing_2011, teather_pointing_2013, barrera_machuca_effect_2019, batmaz_effect_2019}. Because of the ISO standard selection task's overwhelming acceptance in the research community, we used it in our preliminary study.

After conducting the preliminary experiment, however, some shortcomings of the ISO standard task became apparent. First, the layout of the targets in the selection task is very limiting. Because the targets are arranged in a circular pattern, it is difficult to test conditions with low $\alpha$ and high $\omega$ values, since this results in targets that significantly overlap one another. This is an issue because many selections in practical 3D user interfaces have exactly these properties---a short distance to move the pointing ray and a large target (e.g., selections on a floating menu).

Second, the trials in the standard selection task are interdependent, because the $\alpha$ and $\omega$ values cannot be altered once a user begins a given selection trial. This is an issue because users could develop muscle memory when repeatedly selecting the targets which would adversely affect the collected data.

Third, the measured time to complete a trial in the ISO standard task is not just a function of the current $\alpha$ and $\omega$ values. Since a trial begins as soon as the previous one ends, the pointing ray may not be pointing at the starting point (it could be anywhere inside the previous target), and the user could even still be moving their hand in the direction of the previous trial, forcing them to slow down, stop, and start moving in the opposite direction for the new trial. To obtain a better model of the time taken for a single pointing ray movement, it is necessary to separate the end of one trial from the beginning of the previous one and to ensure that the pointing ray's initial direction is very close to the starting point for the trial.

Interestingly, we are not the first to call into question the validity and effectiveness of the ISO standard selection task. Batmaz and Stuerzlinger also suggested in their work that it could be time to explore alternate options for a task in distal pointing work due to the age of the ISO standard selection task which has been used for over 20 years in distal pointing work \cite{batmaz_effective_2022}.

\subsection{Updated Distal Pointing Method}
In order to remedy the aforementioned issues with ISO 9241-411, we propose an updated distal pointing methodology that allows for greater flexibility and accuracy in evaluating user distal selection performance. Specifically, our new methodology allows us to study a wide range of combinations of $\alpha$ and $\omega$; it allows these values to be altered for each trial independently, without having to complete an entire set of trials consisting of targets with the same $\alpha$ and $\omega$ values; and it separates the trials temporally so that the end of one trial is not the start of the next. 

This is accomplished through a new layout that features selection target locations spaced equally apart (10\degree {} between each target both vertically and horizontally) on a hemispherical grid. We use several visual components to illustrate the current trial (see Figure \ref{fig:teaser}). Users begin a trial by pointing to a blue starting object with an angular size of 1\degree {} (the small size ensures that the starting location is consistent for all users). Users are directed to this starting object by a blue ray that is attached to the starting object and follows the ray emanating from the user's controller. The starting object also features a white arrow that informs the user of the direction that they will need to go in order to reach the target. Adding the arrow prevents the user from having to search for the target, which will allow for a more accurate measure of pure user performance. It is also worth noting that the starting object and target are always in the participant's field of view. The target is yellow in color and features an angular size of $\omega$. Prior to the time for a trial beginning, participants are shown both the starting object and target which eliminates the need for searching during the trial. Eliminating the need to search ensures that the trials are based solely on movement. Once the user has located the starting object and pointed to it, they click the trigger to begin the trial, then move as quickly as possible to point to any location inside the target and click the trigger again to select it and end the trial. Audio feedback is provided to the user whenever they successfully select either the starting object or target. The visuals of the new distal pointing methodology can be seen in Figure \ref{fig:teaser}. 

This setup allows us to study $\alpha$ values in a range from 10-70\degree {} and $\omega$ values up to 136\degree{}. We are limited to these values due to the field of view (FOV) of the Meta Quest 2 HWD which is estimated to have a 104\degree {} horizontal FOV and a 98\degree {} vertical FOV \footnote{https://risa2000.github.io/hmdgdb/}. It would be possible to add more $\alpha$ and $\omega$ values as the FOV improves on future HWDs. It should also be noted that not all combinations of $\alpha$ and $\omega$ are possible, because $\alpha$ must be at least $\omega$/2 in order to avoid situations where the starting point is inside the target.

\section{Experiment}
\subsection{Goals}
The main goal of the second study was to generate the necessary data to test a variety of distal pointing models when $\alpha$ and $\omega$ were not constrained by the ISO 9241-411 pointing task. Based on the results of the two-part model from the preliminary study (section 3), we wanted to verify if a two part approach could provide a better fit than solely using the $ID_{\text{ANG}}$ model when using a wide range of ID values.  

\subsection{Experimental Design}
Our independent variables were $\alpha$ and $\omega$. There were seven possible values for $\alpha$ ranging from 10-70 degrees in increments of 10 degrees. There were 17 possible values of $\omega$: 2, 8, 16, 24, 32, 40, 48, 56, 64, 72, 80, 88, 96, 104, 112, 120, 128, and 136 degrees. There were only 72 possible combinations of $\alpha$ and $\omega$ due to the constraints described at the end of Section 4.2. We intentionally chose combinations of $\alpha$ and $\omega$ that would provide low ID values since we were especially interested in modeling easy distal pointing tasks. 

The dependent variable was the time it takes a participant to complete a selection task. Trials were timed from the moment when the starting object was selected to the moment when the target was selected. Participants could attempt to select the starting object as many times as necessary. However, if participants did not successfully hit the target, then the trial was discarded and was given to the participant again at the end of the study.

\subsection{Apparatus}
The same equipment (i.e. HWD, PC, etc.) used in the preliminary study was reused for the main study. 
  
\subsection{Procedure}
Much of the procedure remained the same between the preliminary and main study. The main procedural differences between the two were specific to the new distal pointing methodology which featured a new training session dedicated to the new task. Participants completed the same questionnaire from the preliminary study and were introduced to the equipment in the same way.

After all of the setup was completed for the study, participants then began the training phase of the study. Participants were instructed to complete the selection trials as quickly and accurately as possible. During the training phase of the study, participants needed to complete at least 10 trials with varying $\alpha$ and $\omega$ values successfully in order to pass and complete the training phase of the study. After completing training, participants then began their first session of actual trials. Each session had 48 trials. The entire study had 15 sessions for a total of 720 trials for each participant. At the beginning of the study, a random list of pairings of $\alpha$ and $\omega$ were generated for each distinct participant. These pairings were randomly distributed throughout each of the sessions.  Participants removed the headset and took a break in between each of the sessions. Participants repeated any missed trials at the end of the study.   

\subsection{Participants}
We recruited 22 participants (15 male, 7 female) from various Human-Computer Interaction and Computer Science email lists. We excluded data from two of these participants because of unusual circumstances that could have biased their results. Specifically, one participant used both hands to maneuver the controller between targets and the other participant wore an arm sling which adversely affected their ability to properly select the targets. 

This left us with 20 participants total (14 male, 6 female) whose data we analyzed. All 20 participants self-reported as being predominantly right handed. They had an average age of 22.31, with ages ranging from 19 to 35. Our participant pool ranked their experience with VR with an average of 3.05 on a five-point scale. They also self-reported their fatigue level, which averaged 2.05 on a five-point scale. Finally, all of our participants self-identified as students. 
\subsection{Results}
In total, 597 of the 14,997 total collected trials were considered to be error trials which were to be repeated at the end of the study session. After these failed trials were redone, there were 14,400 valid trials that could used for analysis. In addition, any trial with a time value that was two standard deviations below or above the mean was removed prior to analysis. 523 trials were removed in this way as an outlier. This left 13,868 trials to be used for analysis.

\subsubsection{User Selection Behavior}
\begin{figure}[ht]
    \centering
    \includegraphics[width=8cm]{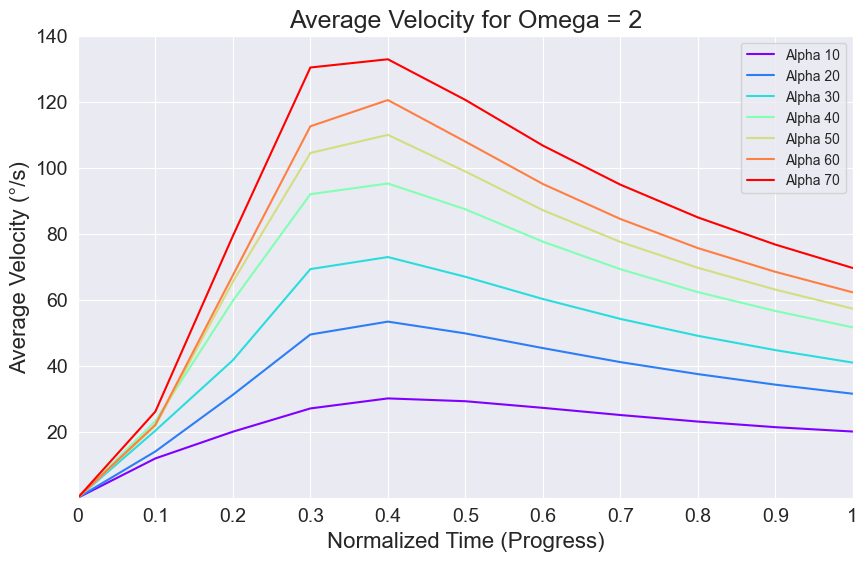}
    \caption{Plot showing the average velocity profile for different values of $\alpha$ when $\omega$ = 2}
    \label{velocity2}
\end{figure}

\begin{figure}[ht]
    \centering
    \includegraphics[width=8cm]{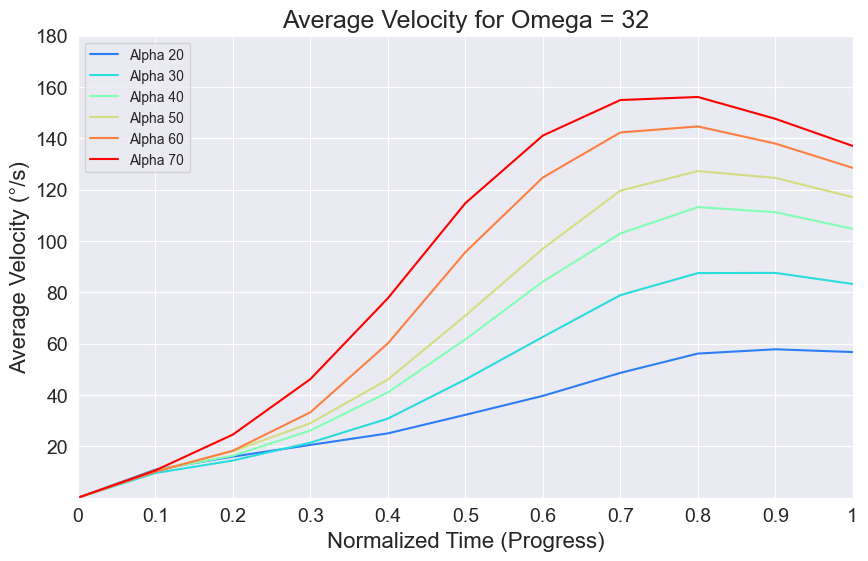}
    \caption{Plot showing the average velocity profile for different values of $\alpha$ when $\omega$ = 32}
    \label{velocity32}
\end{figure}

Distal pointing performance depends on how, mechanically, users manipulate the input device when completing pointing tasks. The standard explanation of user behavior is that aimed movements consist of a high-velocity, less precise ballistic phase, and a low-speed, more precise correction phase\cite{woodworth_accuracy_1899, meyer_models_nodate, meyer_optimality_nodate}. Moreover, standard models assume that users select the center of targets. Based on intuition and the results of the preliminary study, we theorized that, for easy distal pointing tasks with large $\omega$ values, users may exhibit purely ballistic behavior, and may select targets close to their near edge. To examine the evidence for these ideas, we conducted an analysis on the behavior of participants in the main study. 

First, velocity profiles were created for each combination of $\alpha$ and $\omega$ in intervals of 10\% (i.e. 10\% of the trial complete, 20\%, etc.). Each velocity value leading up to and including the velocity at each 10\% interval were averaged together. Note that only the velocity values after the previous interval were included in the averaging process (i.e. values from the 10\% average velocity were not reused in the 20\% average.) Figures \ref{velocity2} and \ref{velocity32} show the velocity profiles when $\omega$ was 2$\degree$ and 32$\degree$, respectively. The plots show that when the $\omega$ value is smaller, participants began the trial and sped up very quickly and began slowing down much sooner compared to Figure \ref{velocity32}. This behavior is really only present when $\omega$ is small (i.e. $\omega$ $\leq$ 24). Otherwise, participants ramp up to ballistic speed, though it takes longer to reach the maximum velocity, and maintain most of this speed through the duration of the trial as shown in Figure \ref{velocity32}. It is also worth noting that in general, larger $\alpha$ values produced larger average velocity values as the starting object and target were far enough away from one another that the participant had ample time to build up speed on their way to the target. However, we also observed that once $\omega$ crossed a threshold ($\omega$ $\geq$ 72) the average velocity values began to decrease. 

To further analyze selection behavior, we explored the distribution of hitpoints within the targets. A hitpoint is defined as the intersection of the ray emanating from the participant's controller with the target at the time of selection. The 3D hitpoint coordinates were projected to 2D to enable visual analysis. 

To obtain the 2D hitpoints, the starting object was first projected onto the plane of the target. The target center was then translated to the origin of the coordinate system. The target, starting object, and hitpoint were then rotated so that the target and other points were aligned with the XY plane. Another rotation was then performed so that the starting point was aligned with the negative x-axis. Doing this rotation ensured that all trial movements were aligned in the same direction. Finally, the points were scaled so that the X value of the hitpoint fell between $-1\leq X \leq1$. In this representation, all trials move from left to right, and all targets have been scaled to have the same size.

\begin{figure*}
  \centering
  \includegraphics[width=\textwidth]{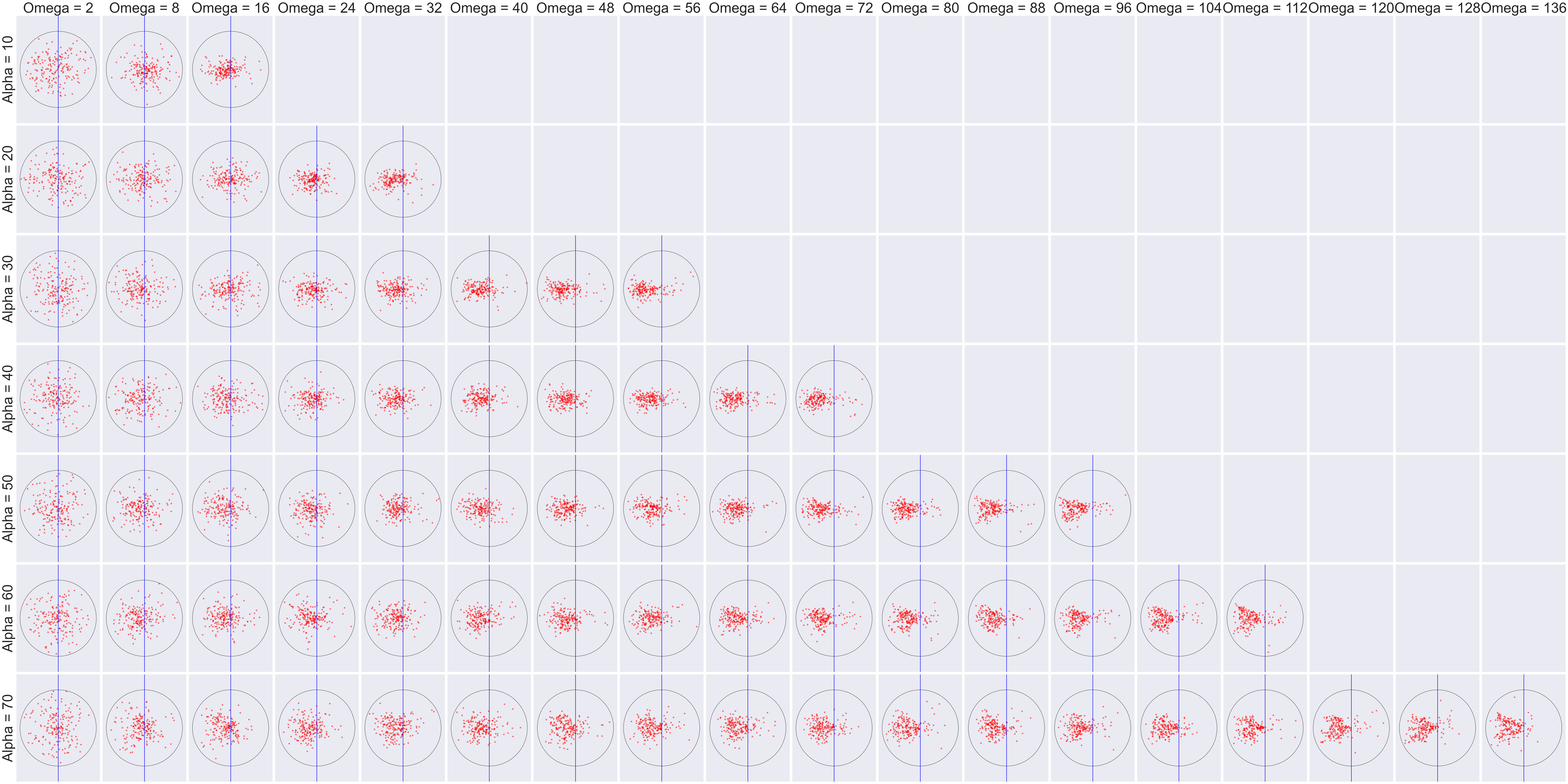}
  \caption{Matrix of plots showing the distribution of hitpoints for each pairing of $\alpha$ and $\omega$}
  \label{scattermatrix}
\end{figure*}

Figure \ref{scattermatrix} shows a matrix of plots that display the distribution of hitpoints for the various combinations of $\alpha$ and $\omega$ in the study. The circle in each plot represents the target, and the line at X = 0 separates hitpoints on the left (undershoot) from those on the right (overshoot). Visual analysis of these plots suggests that trials with smaller $\omega$ values have a more even distribution of hitpoints, while trials with larger $\omega$ values result in more undershoot.

A mixed linear model was created to evaluate the effects of $\alpha$ and $\omega$ on the X-component of the hitpoint coordinate using a linear mixed model in JASP. We found that $\alpha$, $\omega$, and their interaction all had a statistically significant effect on the X coordinate value (p $<$ .0001). The average X-coordinate of the hitpoint decreases as both $\alpha$ and $\omega$ increase.) 

In summary, when looking at user selection behavior, we saw that for large $\omega$ values, users' speed increased as a trial progressed and they maintained most of that speed throughout the trial. Furthermore, we found that as $\omega$ increased, hitpoints increasingly got closer to the edge of the target. These analyses suggest that large targets may have faster performance than expected by standard models due to more purely ballistic selection movement and users taking advantage of large targets to reduce the effective $\alpha$. These findings reinforced our intention to explore a two-part model in which the easy trials are modeled as a function of $\alpha$ only. 

\subsubsection{Model Fitting}
After conducting both the hitpoint and velocity analyses, a new two-part model was created in which the same methodology from the preliminary study was followed with the exception of the IDs used in the split left and right models. The right model still used the respective ID calculation for whatever ID model was being used at the time, however, the left model only utilized the $\alpha$ value in its regression calculations as $\alpha$ has been used in previous literature to model purely ballistic, easy distal pointing tasks \cite{gan_geometrical_1988, Hoffmann_Hui_2010, Lin_Tsai_2015}.

\begin{figure}[ht]
  \centering
  \includegraphics[width=8cm]{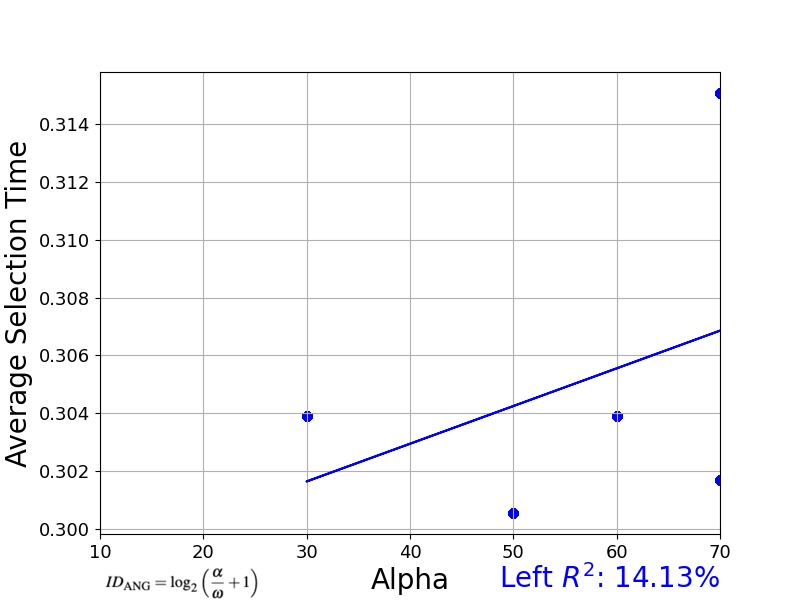}
  \caption{Linear regression plot for the best fitting two-part model using the $ID_{\text{ANG}}$ model and Alpha at breakpoint = 0.6293. This plot shows the left portion of the model (Alpha Only)}
  \label{BestFittingTwoPartModelWithAlphaMainStudyLeft}
\end{figure}

\begin{figure}[ht]
  \centering
  \includegraphics[width=8cm]{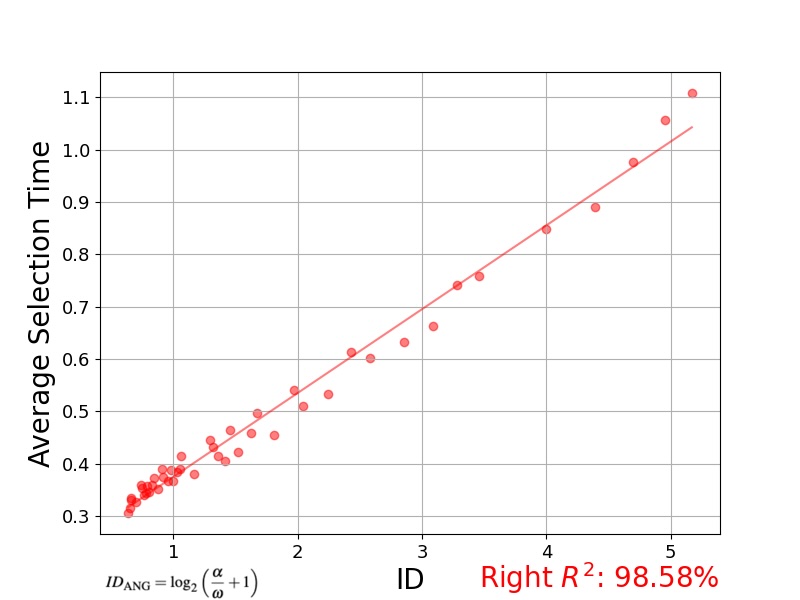}
  \caption{Linear regression plot for the best fitting two-part model using the $ID_{\text{ANG}}$ model and Alpha at breakpoint = 0.6293. This plot shows the right portion of the model ($ID_{\text{ANG}}$)}
  \label{BestFittingTwoPartModelWithAlphaMainStudyRight}
\end{figure}

Despite the evidence from user behavior, we found that easier distal pointing tasks could not be modeled as a linear function of $\alpha$ alone. The left model (the one solely using $\alpha$) had very low \( R^2 \) values (\( R^2 \) $<$ 15\%). We show the most successful two-part model in Figures \ref{BestFittingTwoPartModelWithAlphaMainStudyLeft} and \ref{BestFittingTwoPartModelWithAlphaMainStudyRight}. As the plots shows, the right side of the model (red) that is using the $ID_{\text{ANG}}$ model provides an excellent fit for the data. However, the left model (Alpha only) provides a very poor fit for the data. The slope for the left model is near 0 which indicates that it would be a poor predictor of the dependent variable (Time).

After determining that $\alpha$ alone was not a good predictor of easy distal pointing performance, the performance of the two-part model when using ID models on both the left and right were tested to see if the two-part model methodology would model easy distal pointing tasks better than any one-size-fits-all model. 

Two-part models were created for the  $ID_{\text{ANG}}$, $ID_{\text{ANG}^3}$, and $ID_{\text{DP}}$ formulations. Looking at the $R^2$ values, we determined that the two-part model approach did not provide a good fit when using the $ID_{\text{ANG}^3}$, and $ID_{\text{DP}}$ formulations. It was often the case with both of these models that only the left or right model provided a good fit or both models provided a mediocre fit for the study data. 

\begin{figure}[ht]
  \centering
  \includegraphics[width=8cm]{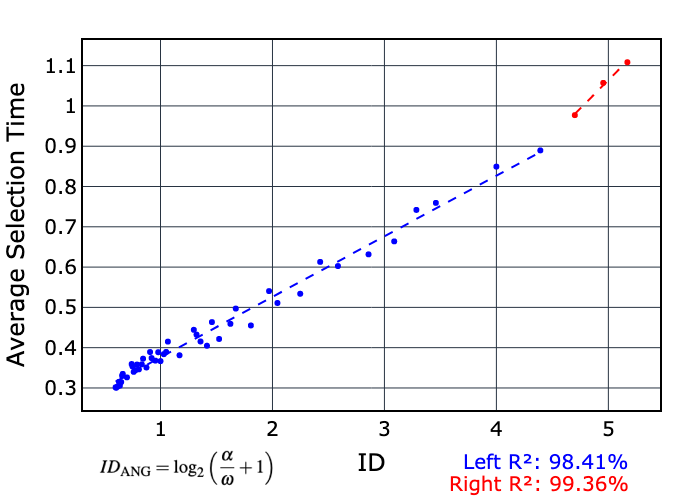}
  \caption{Linear regression plot for the best fitting two-part model using the $ID_{\text{ANG}}$ model at breakpoint = 4.392}
  \label{BestFittingTwoPartModelMainStudy}
\end{figure}

The two-part model for $ID_{\text{ANG}}$ model could provide a high level of fit (98-99\%), but the best fit was when the breakpoint between the left and right models was at a high ID value, which is incongruent with our hypothesis that we should model easy distal pointing tasks separately. In addition, as Figure \ref{BestFittingTwoPartModelMainStudy} shows, even in the best case, the slopes of both the left and right model are very similar to one another, which suggests that there is little benefit to using a two-part model approach. This motivated us to look again at simpler one-part models. 

\begin{figure}[ht]
    \centering
    \includegraphics[width=8cm]{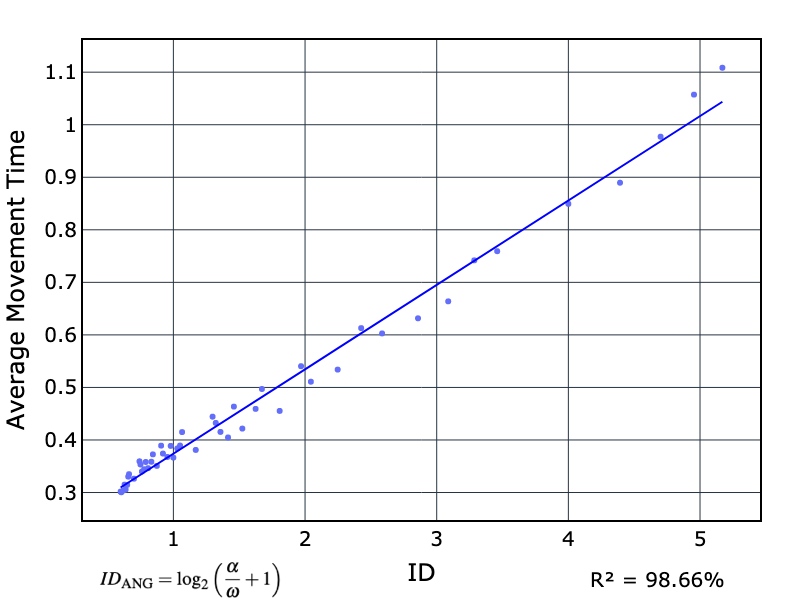}
    \caption{Linear regression plot using the data from the main study for the $ID_{\text{ANG}}$ model. }
    \label{fig:IDANGMainStudy}
\end{figure}

\begin{figure}[ht]
    \centering
    \includegraphics[width=8cm]{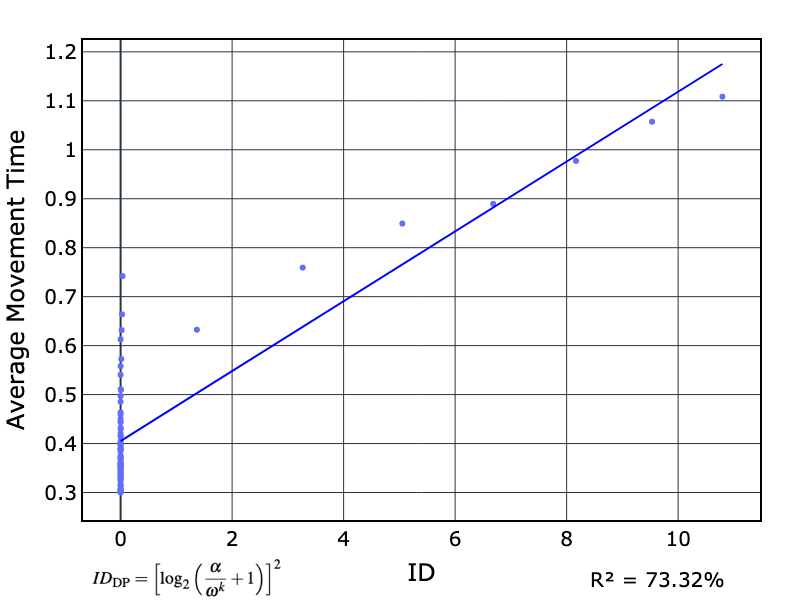}
    \caption{Linear regression plot using the data from the main study for the $ID_{\text{DP}}$ model. }
    \label{fig:IDDPMainStudy}
\end{figure}

Therefore, we fit the data from the main study to the $ID_{\text{ANG}}$, $ID_{\text{ANG}^3}$, and $ID_{\text{DP}}$ models. The models produced \( R^2 \) values of 97.86\%, 77.96\%, and 73.3\% respectively. The linear regression plots for the $ID_{\text{ANG}}$ and $ID_{\text{DP}}$ models are shown in Figure \ref{fig:IDANGMainStudy} and Figure \ref{fig:IDDPMainStudy}. Looking at Figure \ref{fig:IDDPMainStudy}, it is clear that easy distal pointing tasks are still not being modeled well by this model. For brevity, the linear regression plot for $ID_{\text{ANG}^3}$ was not included, however, it tells a very similar story to the $ID_{\text{DP}}$ model and also does not model easy distal pointing tasks well.

As Figure \ref{fig:IDANGMainStudy} and the associated \( R^2 \) value shows, the $ID_{\text{ANG}}$ model produces an excellent fit for the data from the main study. The $ID_{\text{ANG}}$ model produced the highest fit of the three models from the Kopper paper \cite{kopper_human_2010} and is also the most elegant of the three previously discussed models as it is essentially an angular version of Fitts' law.

\section{Discussion}
\subsection{RQ1 - Revisiting \& Updating Fitts' Law Performance Models}
Our first research question asked: 
\textbf{``Is there a simple and elegant model that can accurately predict distal pointing performance across a wide range of realistic task difficulties?''}

Preliminary evidence and intuition led us to believe that we might need to model easy distal pointing tasks separately, perhaps as a function of $\alpha$ alone. However, we found that the two-part model was ultimately unnecessary or unreliable. In creating a two-part model with the $ID_{\text{ANG}}$ formulation, we found a larger \( R^2 \) value on both the left and right split regression models at \( R^2 \) = 98\% and 99\% respectively. However, the gains from the two-part model are ultimately not worth the added complexity due to the similarity of the slopes on the left and right models. In the end, we found that a one-part model based on the $ID_{\text{ANG}}$ formulation produces the highest \( R^2 \) value at 97.86\%. The $ID_{\text{ANG}}$ model is simple, elegant, and easy to understand, as it is essentially the Shannon formulation of Fitts' law with angular measurements instead of linear measurements.

We did not produce similar results to Kopper et al. when utilizing the $ID_{\text{DP}}$ model which was their most compatible and preferred model\cite{kopper_human_2010}. It is possible that because of the newly proposed distal pointing methodology, we were able to generate a greater variety of ID values which allowed us to better test the distal pointing models. In addition, it is possible that the older display and tracking technology used by Kopper et al. could have affected their study results. Finally, Kopper's more complicated model may be simply be a result of overfitting to the data.

\subsection{RQ2 - User Selection Behavior's Effect on Performance Modeling}

User selection behavior varied greatly depending upon the selected value for both $\alpha$ and $\omega$. We asked: \textbf{``How does user selection behavior change based on task difficulty, and how can that inform the design of a predictive model?''}

Our hitpoint analysis showed that larger $\omega$ (and to some extent $\alpha$) values resulted in a greater proportion of undershoot. Users can take advantage of large targets to reduce the total angular distance of the movement. However, given the failure of our two-part ballistic-only model to predict movement time for easy tasks, it does not appear that this reduced angular distance actually made selection faster.

This is partially explained by our velocity analysis. We did indeed find that users slowed their movements less near the end of a trial with larger $\omega$ values. However, we also found that for very large targets ($\omega$ $\geq$ 72), average overall velocity values decrease. We hypothesize that the larger target sizes size influenced the participant's perception of the target and they felt less inclined to quickly select the target because it was so large.

Overall, the evidence suggests that, while distal pointing behavior is not consistent across a wide range of IDs, the effects of extreme $\alpha$ and $\omega$ values can cancel each other out, leading to the simple $ID_{\text{ANG}}$ model being the best predictor of movement time for both very easy and more difficult distal pointing tasks.

\subsection{Evaluation Of Our New Distal Pointing Method}
As we have described, there are numerous practical improvements that our new method provides (i.e., greater range of ID values, lack of interdependence among trials, temporal separation of trials) Moreover, looking at the data collected using our new method, and comparing it to our preliminary data that was collected using the ISO standard task, reveals an advantage that was not immediately obvious at first glance. 

In the preliminary study, we discovered that the regression for the $ID_{\text{ANG}}$ model had a negative intercept of -0.0406 which is undesirable as it does not make sense in the context of the distal pointing prediction model. However, in the data collected using our new method, when we use the same $ID_{\text{ANG}}$ model and run a regression on that data, we find that it has an intercept of 0.2133. Because our new method allowed us to test very easy distal selection tasks (i.e., Those tasks with a very small ID value) and test combinations of $\alpha$ and $\omega$ that otherwise would have been impossible with the ISO standard task, we think that our new method allowed us to generate a greater variety of data which produced a more realistic intercept. This further contributes to our goal of finding a simple, yet explainable performance model. 

\section{Limitations \& Future Work}
This work is not without its limitations. First, our participant pool could have been more diverse in many respects. In particular, all of our participants self-identified as being predominantly right-handed. While this is not inherently a limitation, since we required participants to use their dominant hand, further work should be conducted with left-handed users to ensure similar results are found. Participants were also fairly young with an average age of 22.31. Further work could be conducted with an older population of participants to see how distal pointing behavior changes, if at all.

Future work could also be conducted to see if the direction a participant moves to reach the target or the tilt of the controller affects the accuracy of the predicted movement time. Additionally, we plan to validate both the $ID_{ANG}$ model as well as the newly proposed distal pointing method in a future work. Additionally, because the new distal pointing task featured independent trials, effective width and throughput were not used in this work as they typically show the variance in user selection behavior through multiple, sequential trials with the same $\alpha$ and $\omega$ values\cite{mackenzie2001accuracy}. Future work should explore whether effective width and throughput are still valid metrics when used with the new distal pointing task. Finally, we have only collected data from a single apparatus. We believe that the model should be generally applicable to a variety of head-worn displays and spatial input devices, only requiring changes to the model's coefficients to accurately predict distal pointing time with different hardware. However, we have yet to validate this claim.

\section{Conclusion}
In this paper, we revisited and updated distal selection performance models. We began by conducting a preliminary study that used the current ISO standard task. Using the data from that study, we found that distal pointing tasks with a low ID level were not modeled well. This led us to exploring the idea of a two-part model that modeled easy and regular distal pointing tasks separately. We conducted a second study with a new distal pointing methodology that allowed us to test a greater range of ID values. We found that ultimately, the two-part model was unnecessary and that the  $ID_{\text{ANG}}$ model, essentially an angular version of Fitts' law, provided the greatest overall fit for both the ISO standard task and our new distal pointing task.

\acknowledgments{
This research was funded by a grant from the U.S. Office of Naval Research. We would also like to thank our study participants for their contribution to our work.}

\bibliographystyle{abbrv-doi}

\bibliography{Main}

\begin{thebibliography}{10}

\bibitem{ahlstrom2005modeling}
D.~Ahlstr{\"o}m.
\newblock Modeling and improving selection in cascading pull-down menus using fitts' law, the steering law and force fields.
\newblock In {\em Proceedings of the SIGCHI conference on Human factors in computing systems}, pp. 61--70, 2005.

\bibitem{amini2025review3dfitts}
M.~Amini, W.~Stuerzlinger, R.~J. Teather, and A.~U. Batmaz.
\newblock A systematic review of fitts' law in 3d extended reality.
\newblock In {\em Conference on Human Factors in Computing Systems}, CHI '25, Apr 2025.
\newblock To appear. doi: {{%
10\hspace{.1pt}\discretionary{.}{%
}{.}\hspace{.4pt}1145\discretionary{/}{%
}{/}\discretionary{/}{%
}{/}3706598\hspace{.1pt}\discretionary{.}{%
}{.}\hspace{.4pt}3713623}}


\bibitem{barrera_machuca_effect_2019}
M.~D. Barrera~Machuca and W.~Stuerzlinger.
\newblock The {Effect} of {Stereo} {Display} {Deficiencies} on {Virtual} {Hand} {Pointing}.
\newblock In {\em Proceedings of the 2019 {CHI} {Conference} on {Human} {Factors} in {Computing} {Systems}}, pp. 1--14. ACM, Glasgow Scotland Uk, May 2019. doi: {{%
10\hspace{.1pt}\discretionary{.}{%
}{.}\hspace{.4pt}1145\discretionary{/}{%
}{/}3290605\hspace{.1pt}\discretionary{.}{%
}{.}\hspace{.4pt}3300437}}


\bibitem{batmaz_effect_2019}
A.~U. Batmaz and W.~Stuerzlinger.
\newblock The {Effect} of {Rotational} {Jitter} on {3D} {Pointing} {Tasks}.
\newblock In {\em Extended {Abstracts} of the 2019 {CHI} {Conference} on {Human} {Factors} in {Computing} {Systems}}, pp. 1--6. ACM, Glasgow Scotland Uk, May 2019. doi: {{%
10\hspace{.1pt}\discretionary{.}{%
}{.}\hspace{.4pt}1145\discretionary{/}{%
}{/}3290607\hspace{.1pt}\discretionary{.}{%
}{.}\hspace{.4pt}3312752}}


\bibitem{batmaz_effective_2022}
A.~U. Batmaz and W.~Stuerzlinger.
\newblock Effective {Throughput} {Analysis} of {Different} {Task} {Execution} {Strategies} for {Mid}-{Air} {Fitts}' {Tasks} in {Virtual} {Reality}.
\newblock {\em IEEE Transactions on Visualization and Computer Graphics}, 28(11):3939--3947, Nov. 2022. doi: {{%
10\hspace{.1pt}\discretionary{.}{%
}{.}\hspace{.4pt}1109\discretionary{/}{%
}{/}TVCG\hspace{.1pt}\discretionary{.}{%
}{.}\hspace{.4pt}2022\hspace{.1pt}\discretionary{.}{%
}{.}\hspace{.4pt}3203105}}


\bibitem{burno_applying_2015}
R.~A. Burno, B.~Wu, R.~Doherty, H.~Colett, and R.~Elnaggar.
\newblock Applying {Fitts}’ {Law} to {Gesture} {Based} {Computer} {Interactions}.
\newblock {\em Procedia Manufacturing}, 3:4342--4349, 2015. doi: {{%
10\hspace{.1pt}\discretionary{.}{%
}{.}\hspace{.4pt}1016\discretionary{/}{%
}{/}j\hspace{.1pt}\discretionary{.}{%
}{.}\hspace{.4pt}promfg\hspace{.1pt}\discretionary{.}{%
}{.}\hspace{.4pt}2015\hspace{.1pt}\discretionary{.}{%
}{.}\hspace{.4pt}07\hspace{.1pt}\discretionary{.}{%
}{.}\hspace{.4pt}429}}


\bibitem{cabric_predictive_2021}
F.~Cabric, E.~Dubois, and M.~Serrano.
\newblock A {Predictive} {Performance} {Model} for {Immersive} {Interactions} in {Mixed} {Reality}.
\newblock In {\em 2021 {IEEE} {International} {Symposium} on {Mixed} and {Augmented} {Reality} ({ISMAR})}, pp. 202--210. IEEE, Bari, Italy, Oct. 2021. doi: {{%
10\hspace{.1pt}\discretionary{.}{%
}{.}\hspace{.4pt}1109\discretionary{/}{%
}{/}ISMAR52148\hspace{.1pt}\discretionary{.}{%
}{.}\hspace{.4pt}2021\hspace{.1pt}\discretionary{.}{%
}{.}\hspace{.4pt}00035}}


\bibitem{cha_extended_2013}
Y.~Cha and R.~Myung.
\newblock Extended {Fitts}' law for {3D} pointing tasks using {3D} target arrangements.
\newblock {\em International Journal of Industrial Ergonomics}, 43(4):350--355, July 2013. doi: {{%
10\hspace{.1pt}\discretionary{.}{%
}{.}\hspace{.4pt}1016\discretionary{/}{%
}{/}j\hspace{.1pt}\discretionary{.}{%
}{.}\hspace{.4pt}ergon\hspace{.1pt}\discretionary{.}{%
}{.}\hspace{.4pt}2013\hspace{.1pt}\discretionary{.}{%
}{.}\hspace{.4pt}05\hspace{.1pt}\discretionary{.}{%
}{.}\hspace{.4pt}005}}


\bibitem{clark_extending_2020}
L.~D. Clark, A.~B. Bhagat, and S.~L. Riggs.
\newblock Extending {Fitts}’ law in three-dimensional virtual environments with current low-cost virtual reality technology.
\newblock {\em International Journal of Human-Computer Studies}, 139:102413, July 2020. doi: {{%
10\hspace{.1pt}\discretionary{.}{%
}{.}\hspace{.4pt}1016\discretionary{/}{%
}{/}j\hspace{.1pt}\discretionary{.}{%
}{.}\hspace{.4pt}ijhcs\hspace{.1pt}\discretionary{.}{%
}{.}\hspace{.4pt}2020\hspace{.1pt}\discretionary{.}{%
}{.}\hspace{.4pt}102413}}


\bibitem{DouglasISO}
S.~A. Douglas, A.~E. Kirkpatrick, and I.~S. MacKenzie.
\newblock Testing pointing device performance and user assessment with the iso 9241, part 9 standard.
\newblock In {\em Proceedings of the SIGCHI Conference on Human Factors in Computing Systems}, CHI '99, p. 215–222. Association for Computing Machinery, New York, NY, USA, 1999. doi: {{%
10\hspace{.1pt}\discretionary{.}{%
}{.}\hspace{.4pt}1145\discretionary{/}{%
}{/}302979\hspace{.1pt}\discretionary{.}{%
}{.}\hspace{.4pt}303042}}


\bibitem{fitts_information_1954}
P.~M. Fitts.
\newblock The information capacity of the human motor system in controlling the amplitude of movement.
\newblock {\em Journal of experimental psychology}, 47(6):381, 1954.
\newblock Publisher: American Psychological Association.

\bibitem{gan_geometrical_1988-2}
K.-C. Gan and E.~R. Hoffmann.
\newblock Geometrical conditions for ballistic and visually controlled movements.
\newblock {\em Ergonomics}, 31(5):829--839, May 1988.
\newblock Number: 5. doi: {{%
10\hspace{.1pt}\discretionary{.}{%
}{.}\hspace{.4pt}1080\discretionary{/}{%
}{/}00140138808966724}}


\bibitem{gan_geometrical_1988}
K.-C. Gan and E.~R. Hoffmann.
\newblock Geometrical conditions for ballistic and visually controlled movements.
\newblock {\em Ergonomics}, 31(5):829--839, May 1988. doi: {{%
10\hspace{.1pt}\discretionary{.}{%
}{.}\hspace{.4pt}1080\discretionary{/}{%
}{/}00140138808966724}}


\bibitem{ghasemi_evaluating_2023}
Y.~Ghasemi, H.~Jeong, K.-B. Park, S.~H. Choi, and J.~Y. Lee.
\newblock Evaluating {User} {Interactions} in {Wearable} {Extended} {Reality}: {Modeling}, {Online} {Remote} {Survey}, and {In}-{Lab} {Experimental} {Methods}.
\newblock {\em IEEE Access}, 11:77856--77872, 2023. doi: {{%
10\hspace{.1pt}\discretionary{.}{%
}{.}\hspace{.4pt}1109\discretionary{/}{%
}{/}ACCESS\hspace{.1pt}\discretionary{.}{%
}{.}\hspace{.4pt}2023\hspace{.1pt}\discretionary{.}{%
}{.}\hspace{.4pt}3298598}}


\bibitem{inproceedings_Grossman}
T.~Grossman and R.~Balakrishnan.
\newblock Pointing at trivariate targets in 3d environments.
\newblock pp. 447--454, 04 2004. doi: {{%
10\hspace{.1pt}\discretionary{.}{%
}{.}\hspace{.4pt}1145\discretionary{/}{%
}{/}985692\hspace{.1pt}\discretionary{.}{%
}{.}\hspace{.4pt}985749}}


\bibitem{Hoffmann_Hui_2010}
E.~R. Hoffmann and M.~C. Hui.
\newblock Movement times of different arm components.
\newblock {\em Ergonomics}, 53(8):979–993, July 2010. doi: {{%
10\hspace{.1pt}\discretionary{.}{%
}{.}\hspace{.4pt}1080\discretionary{/}{%
}{/}00140139\hspace{.1pt}\discretionary{.}{%
}{.}\hspace{.4pt}2010\hspace{.1pt}\discretionary{.}{%
}{.}\hspace{.4pt}500403}}


\bibitem{holmes_using_2016}
D.~E. Holmes, D.~K. Charles, P.~J. Morrow, S.~McClean, and S.~M. McDonough.
\newblock Using {Fitt}'s {Law} to {Model} {Arm} {Motion} {Tracked} in {3D} by a {Leap} {Motion} {Controller} for {Virtual} {Reality} {Upper} {Arm} {Stroke} {Rehabilitation}.
\newblock In {\em 2016 {IEEE} 29th {International} {Symposium} on {Computer}-{Based} {Medical} {Systems} ({CBMS})}, pp. 335--336. IEEE, Belfast and Dublin, Ireland, June 2016. doi: {{%
10\hspace{.1pt}\discretionary{.}{%
}{.}\hspace{.4pt}1109\discretionary{/}{%
}{/}CBMS\hspace{.1pt}\discretionary{.}{%
}{.}\hspace{.4pt}2016\hspace{.1pt}\discretionary{.}{%
}{.}\hspace{.4pt}41}}


\bibitem{ISO_2010}
{Ergonomics of human-system interaction}.
\newblock Standard, International Organization for Standardization, Mar. 2010.

\bibitem{JanzenTwoPart}
I.~Janzen, V.~K. Rajendran, and K.~S. Booth.
\newblock Modeling the impact of depth on pointing performance.
\newblock In {\em Proceedings of the 2016 CHI Conference on Human Factors in Computing Systems}, CHI '16, p. 188–199. Association for Computing Machinery, New York, NY, USA, 2016. doi: {{%
10\hspace{.1pt}\discretionary{.}{%
}{.}\hspace{.4pt}1145\discretionary{/}{%
}{/}2858036\hspace{.1pt}\discretionary{.}{%
}{.}\hspace{.4pt}2858244}}


\bibitem{john1995goms}
B.~John.
\newblock Why goms?
\newblock {\em interactions}, 2(4):80--89, 1995.

\bibitem{john1996goms}
B.~E. John and D.~E. Kieras.
\newblock The goms family of user interface analysis techniques: Comparison and contrast.
\newblock {\em ACM Transactions on Computer-Human Interaction (TOCHI)}, 3(4):320--351, 1996.

\bibitem{john1996using}
B.~E. John and D.~E. Kieras.
\newblock Using goms for user interface design and evaluation: Which technique?
\newblock {\em ACM Transactions on Computer-Human Interaction (TOCHI)}, 3(4):287--319, 1996.

\bibitem{kabbash1995prince}
P.~Kabbash and W.~A. Buxton.
\newblock The “prince” technique: Fitts' law and selection using area cursors.
\newblock In {\em Proceedings of the SIGCHI conference on Human factors in computing systems}, pp. 273--279, 1995.

\bibitem{kieras1997guide}
D.~Kieras.
\newblock A guide to goms model usability evaluation using ngomsl.
\newblock In {\em Handbook of human-computer interaction}, pp. 733--766. Elsevier, 1997.

\bibitem{kopper_rapid_2011}
R.~Kopper, F.~Bacim, and D.~A. Bowman.
\newblock Rapid and accurate {3D} selection by progressive refinement.
\newblock In {\em 2011 {IEEE} {Symposium} on {3D} {User} {Interfaces} ({3DUI})}, pp. 67--74. IEEE, Singapore, Singapore, Mar. 2011. doi: {{%
10\hspace{.1pt}\discretionary{.}{%
}{.}\hspace{.4pt}1109\discretionary{/}{%
}{/}3DUI\hspace{.1pt}\discretionary{.}{%
}{.}\hspace{.4pt}2011\hspace{.1pt}\discretionary{.}{%
}{.}\hspace{.4pt}5759219}}


\bibitem{kopper_human_2010}
R.~Kopper, D.~A. Bowman, M.~G. Silva, and R.~P. McMahan.
\newblock A human motor behavior model for distal pointing tasks.
\newblock {\em International Journal of Human-Computer Studies}, 68(10):603--615, Oct. 2010.
\newblock Number: 10. doi: {{%
10\hspace{.1pt}\discretionary{.}{%
}{.}\hspace{.4pt}1016\discretionary{/}{%
}{/}j\hspace{.1pt}\discretionary{.}{%
}{.}\hspace{.4pt}ijhcs\hspace{.1pt}\discretionary{.}{%
}{.}\hspace{.4pt}2010\hspace{.1pt}\discretionary{.}{%
}{.}\hspace{.4pt}05\hspace{.1pt}\discretionary{.}{%
}{.}\hspace{.4pt}001}}


\bibitem{Lin_Tsai_2015}
R.~F. Lin and Y.-C. Tsai.
\newblock The use of ballistic movement as an additional method to assess performance of computer mice.
\newblock {\em International Journal of Industrial Ergonomics}, 45:71–81, Feb. 2015. doi: {{%
10\hspace{.1pt}\discretionary{.}{%
}{.}\hspace{.4pt}1016\discretionary{/}{%
}{/}j\hspace{.1pt}\discretionary{.}{%
}{.}\hspace{.4pt}ergon\hspace{.1pt}\discretionary{.}{%
}{.}\hspace{.4pt}2014\hspace{.1pt}\discretionary{.}{%
}{.}\hspace{.4pt}12\hspace{.1pt}\discretionary{.}{%
}{.}\hspace{.4pt}003}}


\bibitem{mackenzie_note_1989}
I.~S. MacKenzie.
\newblock A {Note} on the {Information}-{Theoretic} {Basis} for {Fitts}’ {Law}.
\newblock {\em Journal of Motor Behavior}, 21(3):323--330, Sept. 1989. doi: {{%
10\hspace{.1pt}\discretionary{.}{%
}{.}\hspace{.4pt}1080\discretionary{/}{%
}{/}00222895\hspace{.1pt}\discretionary{.}{%
}{.}\hspace{.4pt}1989\hspace{.1pt}\discretionary{.}{%
}{.}\hspace{.4pt}10735486}}


\bibitem{mackenzie_fitts_1992}
I.~S. MacKenzie.
\newblock Fitts' {Law} as a {Research} and {Design} {Tool} in {Human}-{Computer} {Interaction}.
\newblock {\em Human–Computer Interaction}, 7(1):91--139, Mar. 1992. doi: {{%
10\hspace{.1pt}\discretionary{.}{%
}{.}\hspace{.4pt}1207\discretionary{/}{%
}{/}s15327051hci0701\_3}}


\bibitem{mackenzie_comparison_nodate}
I.~S. Mackenzie and W.~Buxton.
\newblock A {COMPARISON} {OF} {INPUT} {DEVICES} {IN} {ELEMENTAL} {POINTING} {AND} {DRAGGING} {TASKS}.

\bibitem{mackenzie_extending_1992}
I.~S. MacKenzie and W.~Buxton.
\newblock Extending {Fitts}' law to two-dimensional tasks.
\newblock In {\em Proceedings of the {SIGCHI} conference on {Human} factors in computing systems - {CHI} '92}, pp. 219--226. ACM Press, Monterey, California, United States, 1992. doi: {{%
10\hspace{.1pt}\discretionary{.}{%
}{.}\hspace{.4pt}1145\discretionary{/}{%
}{/}142750\hspace{.1pt}\discretionary{.}{%
}{.}\hspace{.4pt}142794}}


\bibitem{mackenzie2001accuracy}
I.~S. MacKenzie, T.~Kauppinen, and M.~Silfverberg.
\newblock Accuracy measures for evaluating computer pointing devices.
\newblock In {\em Proceedings of the SIGCHI conference on Human factors in computing systems}, pp. 9--16, 2001.

\bibitem{meyer_optimality_nodate}
D.~E. Meyer, S.~Kornblum, R.~A. Abrams, and C.~E. Wright.
\newblock Optimality in {Human} {Motor} {Performance}: {Ideal} {Control} of {Rapid} {Aimed} {Movements}.

\bibitem{meyer_models_nodate}
D.~E. Meyer, J.~E.~K. Smith, C.~E. Wright, B.~Laboratories, and M.~Hill.
\newblock Models for the {Speed} and {Accuracy} of {Aimed} {Movements}.

\bibitem{murata_empirical_1996}
A.~Murata.
\newblock Empirical evaluation of performance models of pointing accuracy and speed with a {PC} mouse.
\newblock {\em International Journal of Human-Computer Interaction}, 8(4):457--469, Oct. 1996. doi: {{%
10\hspace{.1pt}\discretionary{.}{%
}{.}\hspace{.4pt}1080\discretionary{/}{%
}{/}10447319609526164}}


\bibitem{murata_extending_1999}
A.~Murata.
\newblock Extending {Effective} {Target} {Width} in {Fitts}' {Law} to a {Two}-{Dimensional} {Pointing} {Task}.
\newblock {\em International Journal of Human-Computer Interaction}, 11(2):137--152, June 1999. doi: {{%
10\hspace{.1pt}\discretionary{.}{%
}{.}\hspace{.4pt}1207\discretionary{/}{%
}{/}S153275901102\_4}}


\bibitem{murata_extending_2001}
A.~Murata and H.~Iwase.
\newblock Extending {Fitts}' law to a three-dimensional pointing task.
\newblock {\em Human Movement Science}, 20(6):791--805, Dec. 2001. doi: {{%
10\hspace{.1pt}\discretionary{.}{%
}{.}\hspace{.4pt}1016\discretionary{/}{%
}{/}S0167\discretionary{%
}{-}{-}9457\discretionary{%
}{(}{(}01\discretionary{)}{%
}{)}00058\discretionary{%
}{-}{-}6}}


\bibitem{rousson_-square_2007}
V.~Rousson and N.~F. Goşoniu.
\newblock An -square coefficient based on final prediction error.
\newblock {\em Statistical Methodology}, 4(3):331--340, July 2007. doi: {{%
10\hspace{.1pt}\discretionary{.}{%
}{.}\hspace{.4pt}1016\discretionary{/}{%
}{/}j\hspace{.1pt}\discretionary{.}{%
}{.}\hspace{.4pt}stamet\hspace{.1pt}\discretionary{.}{%
}{.}\hspace{.4pt}2006\hspace{.1pt}\discretionary{.}{%
}{.}\hspace{.4pt}11\hspace{.1pt}\discretionary{.}{%
}{.}\hspace{.4pt}004}}


\bibitem{schmitt1999calculation}
A.~Schmitt and P.~Oel.
\newblock Calculation of totally optimized button configurations using fitts’ law.
\newblock {\em Optimization}, 1:1, 1999.

\bibitem{schuetz_explanation_2019}
I.~Schuetz, T.~S. Murdison, K.~J. MacKenzie, and M.~Zannoli.
\newblock An {Explanation} of {Fitts}' {Law}-like {Performance} in {Gaze}-{Based} {Selection} {Tasks} {Using} a {Psychophysics} {Approach}.
\newblock In {\em Proceedings of the 2019 {CHI} {Conference} on {Human} {Factors} in {Computing} {Systems}}, pp. 1--13. ACM, Glasgow Scotland Uk, May 2019. doi: {{%
10\hspace{.1pt}\discretionary{.}{%
}{.}\hspace{.4pt}1145\discretionary{/}{%
}{/}3290605\hspace{.1pt}\discretionary{.}{%
}{.}\hspace{.4pt}3300765}}


\bibitem{shoemaker}
G.~Shoemaker, T.~Tsukitani, Y.~Kitamura, and K.~S. Booth.
\newblock Two-part models capture the impact of gain on pointing performance.
\newblock {\em ACM Trans. Comput.-Hum. Interact.}, 19(4), dec 2012. doi: {{%
10\hspace{.1pt}\discretionary{.}{%
}{.}\hspace{.4pt}1145\discretionary{/}{%
}{/}2395131\hspace{.1pt}\discretionary{.}{%
}{.}\hspace{.4pt}2395135}}


\bibitem{Sindhupathiraja_2024}
S.~R. Sindhupathiraja, A.~K. M.~A. Ullah, W.~Delamare, and K.~Hasan.
\newblock Exploring bi-manual teleportation in virtual reality.
\newblock In {\em 2024 IEEE Conference Virtual Reality and 3D User Interfaces (VR)}, p. 754–764. IEEE, Mar. 2024. doi: {{%
10\hspace{.1pt}\discretionary{.}{%
}{.}\hspace{.4pt}1109\discretionary{/}{%
}{/}vr58804\hspace{.1pt}\discretionary{.}{%
}{.}\hspace{.4pt}2024\hspace{.1pt}\discretionary{.}{%
}{.}\hspace{.4pt}00095}}


\bibitem{teather_pointing_2011}
R.~J. Teather and W.~Stuerzlinger.
\newblock Pointing at {3D} targets in a stereo head-tracked virtual environment.
\newblock In {\em 2011 {IEEE} {Symposium} on {3D} {User} {Interfaces} ({3DUI})}, pp. 87--94. IEEE, Singapore, Singapore, Mar. 2011. doi: {{%
10\hspace{.1pt}\discretionary{.}{%
}{.}\hspace{.4pt}1109\discretionary{/}{%
}{/}3DUI\hspace{.1pt}\discretionary{.}{%
}{.}\hspace{.4pt}2011\hspace{.1pt}\discretionary{.}{%
}{.}\hspace{.4pt}5759222}}


\bibitem{teather_pointing_2013}
R.~J. Teather and W.~Stuerzlinger.
\newblock Pointing at 3d target projections with one-eyed and stereo cursors.
\newblock In {\em Proceedings of the {SIGCHI} {Conference} on {Human} {Factors} in {Computing} {Systems}}, pp. 159--168. ACM, Paris France, Apr. 2013. doi: {{%
10\hspace{.1pt}\discretionary{.}{%
}{.}\hspace{.4pt}1145\discretionary{/}{%
}{/}2470654\hspace{.1pt}\discretionary{.}{%
}{.}\hspace{.4pt}2470677}}


\bibitem{thompson2004gain}
S.~Thompson, J.~Slocum, and M.~Bohan.
\newblock Gain and angle of approach effects on cursor-positioning time with a mouse in consideration of fitts' law.
\newblock In {\em Proceedings of the Human Factors and Ergonomics Society Annual Meeting}, vol.~48, pp. 823--827. SAGE Publications Sage CA: Los Angeles, CA, 2004.

\bibitem{triantafyllidis_challenges_2021}
E.~Triantafyllidis and Z.~Li.
\newblock The {Challenges} in {Modeling} {Human} {Performance} in {3D} {Space} with {Fitts}’ {Law}.
\newblock In {\em Extended {Abstracts} of the 2021 {CHI} {Conference} on {Human} {Factors} in {Computing} {Systems}}, pp. 1--9. ACM, Yokohama Japan, May 2021. doi: {{%
10\hspace{.1pt}\discretionary{.}{%
}{.}\hspace{.4pt}1145\discretionary{/}{%
}{/}3411763\hspace{.1pt}\discretionary{.}{%
}{.}\hspace{.4pt}3443442}}


\bibitem{wagner_fitts_2023}
U.~Wagner, M.~N. Lystbæk, P.~Manakhov, J.~E.~S. Grønbæk, K.~Pfeuffer, and H.~Gellersen.
\newblock A {Fitts}’ {Law} {Study} of {Gaze}-{Hand} {Alignment} for {Selection} in {3D} {User} {Interfaces}.
\newblock In {\em Proceedings of the 2023 {CHI} {Conference} on {Human} {Factors} in {Computing} {Systems}}, pp. 1--15. ACM, Hamburg Germany, Apr. 2023. doi: {{%
10\hspace{.1pt}\discretionary{.}{%
}{.}\hspace{.4pt}1145\discretionary{/}{%
}{/}3544548\hspace{.1pt}\discretionary{.}{%
}{.}\hspace{.4pt}3581423}}


\bibitem{welford1968fundamentals}
A.~T. Welford.
\newblock Fundamentals of skill.
\newblock 1968.

\bibitem{woodworth_accuracy_1899}
R.~S. Woodworth.
\newblock Accuracy of voluntary movement.
\newblock {\em The Psychological Review: Monograph Supplements}, 3(3):i--114, 1899. doi: {{%
10\hspace{.1pt}\discretionary{.}{%
}{.}\hspace{.4pt}1037\discretionary{/}{%
}{/}h0092992}}


\bibitem{zeng_fitts_2012}
X.~Zeng, A.~Hedge, and F.~Guimbretiere.
\newblock Fitts’ {Law} in {3D} {Space} with {Coordinated} {Hand} {Movements}.
\newblock {\em Proceedings of the Human Factors and Ergonomics Society Annual Meeting}, 56(1):990--994, Sept. 2012. doi: {{%
10\hspace{.1pt}\discretionary{.}{%
}{.}\hspace{.4pt}1177\discretionary{/}{%
}{/}1071181312561207}}


\bibitem{zhao_movement_2023}
C.~Zhao, K.~W. Li, and L.~Peng.
\newblock Movement {Time} for {Pointing} {Tasks} in {Real} and {Augmented} {Reality} {Environments}.
\newblock {\em Applied Sciences}, 13(2):788, Jan. 2023.
\newblock Number: 2. doi: {{%
10\hspace{.1pt}\discretionary{.}{%
}{.}\hspace{.4pt}3390\discretionary{/}{%
}{/}app13020788}}


\bibitem{zhou_h-goms_2023}
X.~Zhou, F.~Teng, X.~Du, J.~Li, M.~Jin, and C.~Xue.
\newblock H-{GOMS}: a model for evaluating a virtual-hand interaction system in virtual environments.
\newblock {\em Virtual Reality}, 27(2):497--522, June 2023. doi: {{%
10\hspace{.1pt}\discretionary{.}{%
}{.}\hspace{.4pt}1007\discretionary{/}{%
}{/}s10055\discretionary{%
}{-}{-}022\discretionary{%
}{-}{-}00674\discretionary{%
}{-}{-}y}}


\end{thebibliography}

\begin{tabular}{ m{0.2\textwidth} m{0.75\textwidth} }
    \begin{minipage}{0.15\textwidth}
        \includegraphics[width=\linewidth]{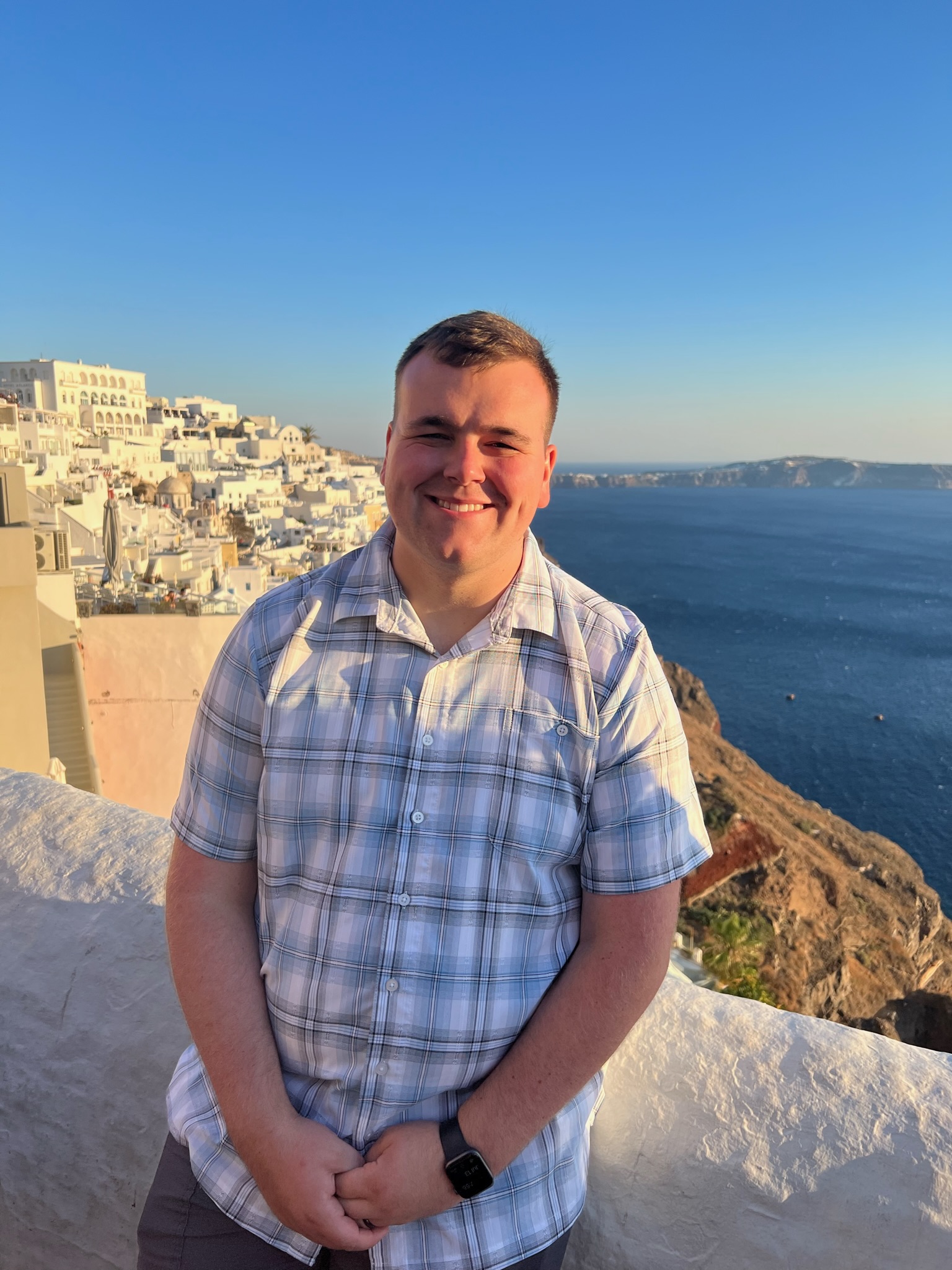}
    \end{minipage} &
    \begin{minipage}{0.27\textwidth}
        \textbf{Logan Lane} \newline Logan Lane is a computer science Ph.D. student at Virginia Tech. He is a member of the 3D Interaction Group and Center for Human-Computer Interaction at Virginia Tech. His research interests include using collaborative, after-action reviews in virtual reality for training in sports as well as using virtual and augmented reality for general collaboration work. \\
    \end{minipage} \\
    \begin{minipage}{0.15\textwidth}
        \includegraphics[width=\linewidth]{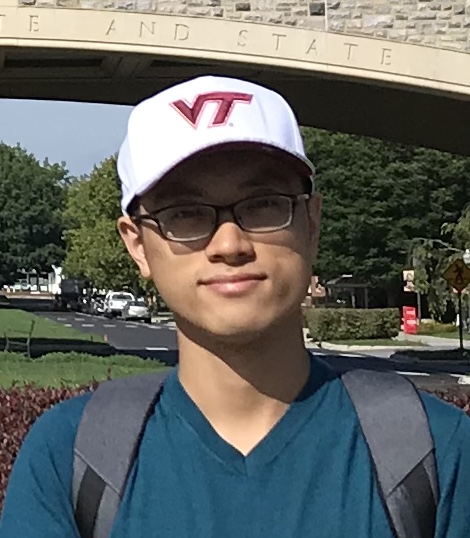}
    \end{minipage} &
    \begin{minipage}{0.27\textwidth}
        \textbf{Feiyu Lu} \newline Feiyu Lu obtained his Ph.D. from Virginia Tech in May 2023. His research interests lie broadly in the intersections of AR/VR, 3DUI, and HCI. His Ph.D. work focuses on enabling lightweight and unobtrusive information display and interactions on AR HWDs to support a variety of everyday tasks. \\
    \end{minipage} \\
    \begin{minipage}{0.15\textwidth}
        \includegraphics[width=\linewidth]{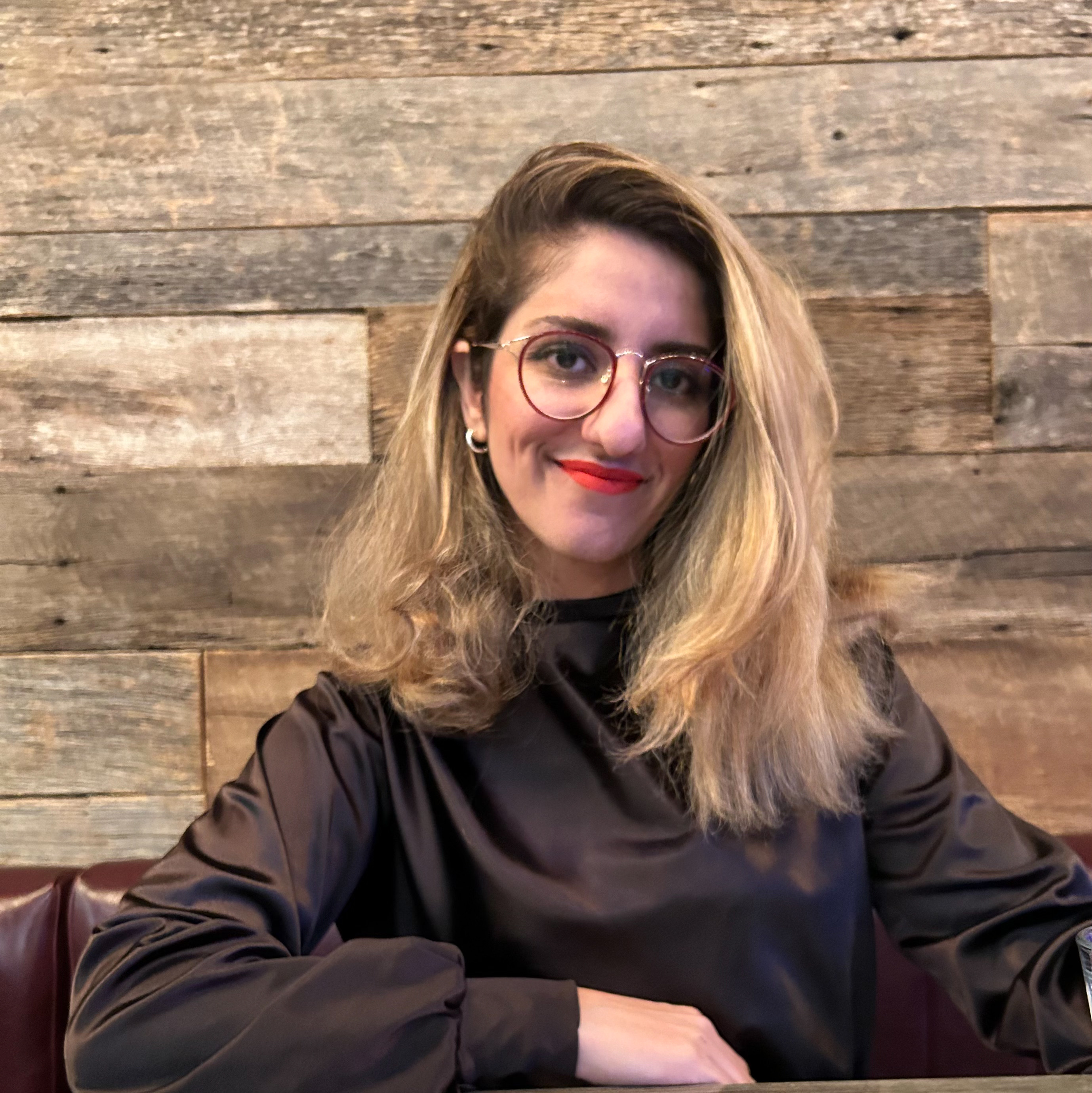}
    \end{minipage} &
    \begin{minipage}{0.27\textwidth}
        \textbf{Shakiba Davari} \newline Shakiba Davari is currently a Ph.D. at the 3D Interaction Lab at Virginia Tech with a focus on intelligent XR interfaces. Throughout her Ph.D., she has contributed methodological guidelines and frameworks for developing intelligent XR, and investigated context-aware AR interfaces that address AR challenges such as real-world occlusion and social intrusiveness.\\
    \end{minipage} \\
    \begin{minipage}{0.15\textwidth}
        \includegraphics[width=\linewidth]{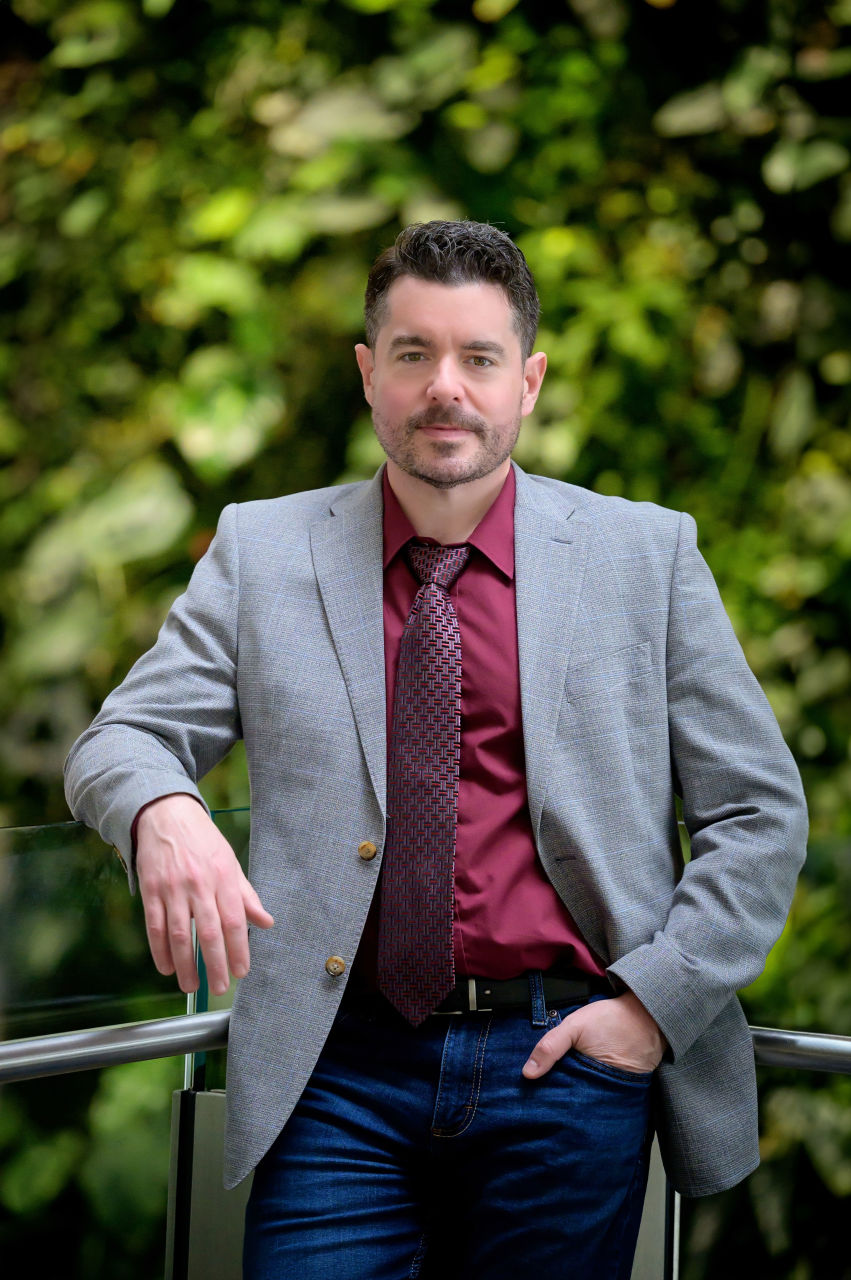}
    \end{minipage} &
    \begin{minipage}{0.27\textwidth}
        \textbf{Robert J. Teather} \newline Robert J. Teather is an Associate Professor in and Director of the School of Information Technology at Carleton University. He is co-director of the MARVEL research group, and studies on human performance in mixed, augmented, and virtual reality environments.
    \end{minipage}\\
    \begin{minipage}{0.15\textwidth}
        \includegraphics[width=\linewidth]{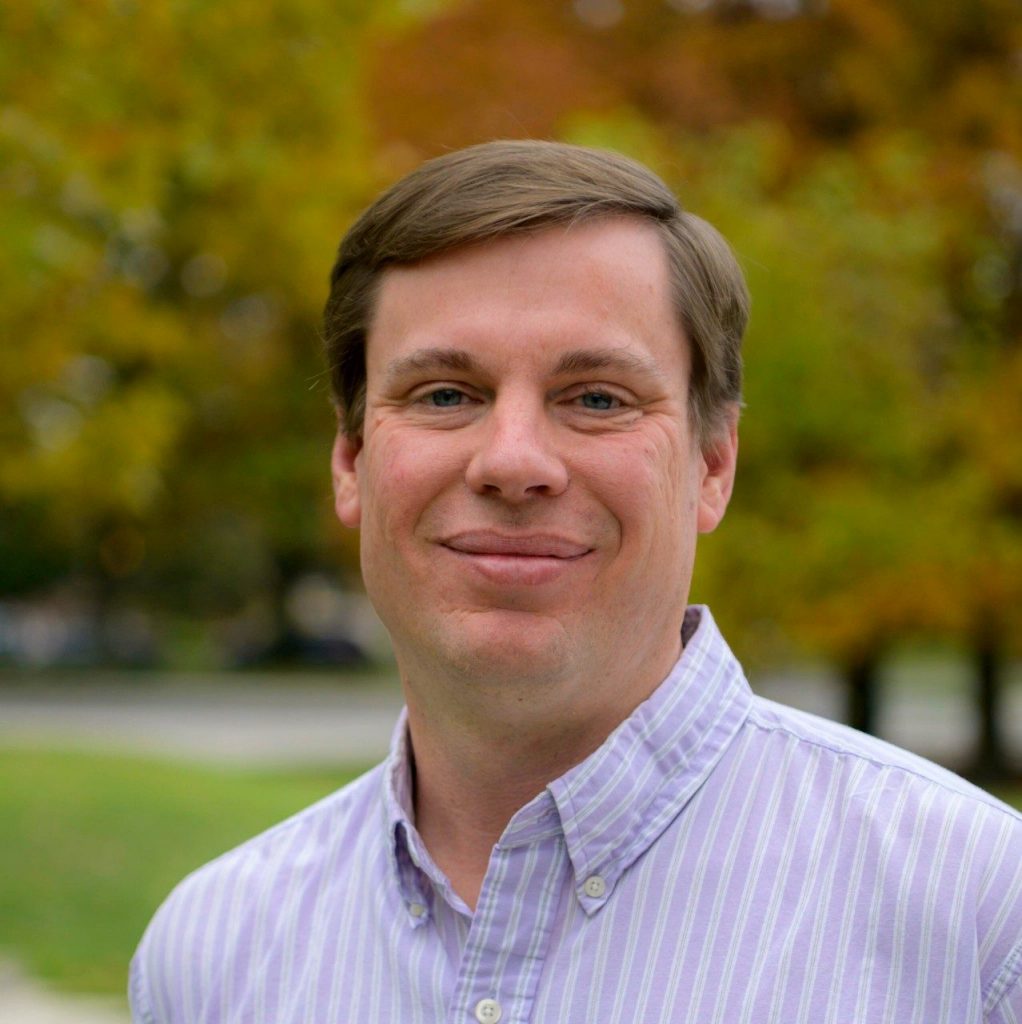}
    \end{minipage} &
    \begin{minipage}{0.27\textwidth}
        \textbf{Doug A. Bowman} \newline Doug A. Bowman is the Frank J. Maher Professor of Computer Science and Director of the Center for Human-Computer Interaction at Virginia Tech. He is the principal investigator of the 3D Interaction Group, focusing on the topics of three-dimensional user interfaces, VR/AR user experience, and the benefits of immersion in virtual environments \\
    \end{minipage} \\
    
\end{tabular}
\end{document}